\documentclass[fleqn,10pt, twocolumn]{wlscirep}
\usepackage[utf8]{inputenc}
\usepackage[T1]{fontenc}
\usepackage{amsmath}
\usepackage{amssymb}
\usepackage{physics}
\usepackage{graphics}
\usepackage{booktabs,multirow}
\usepackage{caption}
\title{The quantum imaginary-time control for accelerating the ground-state preparation}

\author[1,2]{Yu-Cheng Chen}
\author[1,*]{Yu-Qin Chen}
\author[2,4$\dagger$]{Alice Hu}
\author[1,3,$\ddagger$]{Chang-Yu Hsieh}
\author[1]{Shengyu Zhang}
\affil[1]{Tencent Quantum Laboratory, Tencent, Shenzhen, Guangdong, 518057, China}
\affil[2]{Department of Mechanical Engineering, City University of Hong Kong, Hong Kong SAR, 999077, China}
\affil[3]{Innovation Institute for Artificial Intelligence in Medicine of Zhejiang University, College of Pharmaceutical Sciences, Zhejiang University, Hangzhou, 310058, China}
\affil[4]{Department of Materials Science and Engineering, City University of Hong Kong, Kowloon, Hong Kong SAR, 999077, China}

\affil[*]{yuqinchen@tencent.com}
\affil[$\dagger$]{alicehu@cityu.edu.hk}
\affil[$\ddagger$]{kimhsieh2@gmail.com}

\begin{abstract}
Quantum computers have been widely speculated to offer significant advantages in obtaining the ground state of difficult Hamiltonian in chemistry and physics. In this work, we first propose a Lyapunov control-inspired strategy to accelerate the well-established imaginary-time method for ground-state preparation.  
We also dig for the source of acceleration of the imaginary-time process under Lyapunov control with theoretical understanding and dynamic process visualization.
To make the method accessible in the noisy intermediate-scale quantum era, we further propose a variational form of the algorithm that could work with shallow quantum circuits. Through numerical experiments on a broad spectrum of realistic models, including molecular systems, 2D Heisenberg models, and Sherrington–Kirkpatrick models, we show that imaginary-time control may substantially accelerate the imaginary time evolution for all systems and even generate orders of magnitude acceleration (suggesting exponential-like acceleration) for challenging molecular Hamiltonians involving small energy gaps as impressive special cases. Finally, with a proper selection of the control Hamiltonian, the new variational quantum algorithm does not incur additional measurement costs compared to the original variational quantum imaginary-time algorithm.
\end{abstract}
\begin{document}

\maketitle
\section*{Introduction}

Quantum computing holds great promise to accelerate essential computational tasks in many fields, such as cryptography, finance, chemistry, material science, and machine learning\cite{Monz2016, Schaden2002, McArdle2020, Biamonte2017}. Particularly, using a quantum computer to solve chemical problems is deemed one of the most promising areas to first witness a practical quantum advantage\cite{McArdle2020} against classical algorithms. For instance, many efforts have been invested in devising efficient algorithms for finding the ground state of molecular Hamiltonians. Some of the major approaches\cite{Bauer2020} include Variational Quantum Eigensolver (VQE)\cite{Aspuru2014,Grimsley2019,Higgott2019,Cerezo2020}, Quantum Phase Estimation (QPE)\cite{Kitaev1995,Miroslav2007,Nathan2016,OBrien2019}, Quantum imaginary time evolution (QITE) \cite{McArdle2019,Gomes2020,Beach2019,Motta2020}, and quantum power iteration method \cite{Seki2021,Kyriienko2020}.

Among these approaches, variational algorithms have attracted much recent attention with their potential to prepare the ground state of a complex Hamiltonian with a shallow quantum circuit in the noisy intermediate-scale quantum (NISQ) era. However, the hybrid quantum-classical optimization loop of the variational algorithms has soon been pointed out to suffer a few prominent technical challenges. More specifically, the classical optimization not only faces a plethora of local minima but also may encounter notorious barren plateau\cite{McClean2018, Grant2019, Cerezo2020, Wang2020, Holmes2021} on the energy landscape. One possible scheme to mitigate this challenge is to simulate the QITE with the time-dependent variational principle. The variational simulation of the QITE avoids a deep quantum circuit and principally alleviates issues with optimizations, such as the barren plateau and local minimums\cite{McArdle2019}.

Apart from the QITE, another common strategy to prepare a ground state by utilizing the time-dependent Schrodinger's function is adiabatic evolution in the real-time domain. Going beyond the adiabatic regime, the theory of optimal quantum control\cite{Kuang2008, Wang2010, Dong2010, Cong2013} provides a general tool and foundation for designing pulses to drive the desired state transitions in a finite time and in the presence of other constraints. Recently, the theory of Lyapunov quantum control has been used, in the context of hybrid quantum-classical algorithms, to solve classical optimization problems\cite{Magann2021}. Under this formulation, one encodes the solutions to an optimization problem as the ground states of a classical spin system. To prepare a ground state, one temporally modulates the structural form of a Hamiltonian, via pulse engineering, in order to achieve the desired state-to-state transition. With the Lyapunov control theory, one can simply use the system's energy as the Lyapunov function to guide pulse engineering. Through our demonstrative examples below, the control-theory-inspired variational methods clearly exhibit faster convergence as well as enhanced robustness against noises compared to the standard VQE. Despite these encouraging instances, real-time quantum control has hardly been regarded as a practical approach to prepare the ground state of strongly correlated many-body systems, because real-time control requires a careful analysis of the controllability of a given setup which is prohibitive to perform for complex systems. Without full controllability, one cannot successfully steer a quantum system toward a target state as illustrated in our numerical example presented later in the text.

Using a temporally modulated Hamiltonian to steer many-body quantum dynamics is not restricted to the real-time domain. In fact, an extensive body of literature proposed quantum simulation methods involving complex time variables. Specifically, there is found that a QITE under the alternating influences of two different Hamiltonians may accelerate the convergence of the ground-state preparation problems\cite{Beach2019}. The combination of interesting observations on the accelerated convergence of an open-loop control of a many-body QITE and the real-time control-based variational methods for shallow quantum circuits provoke more thoughts on whether one can efficiently prepare ground states based on a close-loop control theory for shallow parametrized quantum circuits in the imaginary-time domain.

We propose an imaginary-time Lyapunov control theory for ground-state preparation in this work. First, we theoretically discuss how steered Hamiltonian can speed up the ground state convergence. Then, we show a proper QITC can converge to the ground state faster than a QITE. For small-gap systems\cite{Campo2012} such as examples on molecules reported in this work, we showcase that a QITC can even converge orders of magnitude faster than the standard QITE, suggesting an exponential-like acceleration. Secondly, we discuss the essential differences between the real- and imaginary-time control theory. For the ground-state preparations, the quantum imaginary time control (QITC) generally admits more lenient conditions on selecting control Hamiltonian to facilitate a given state transition. Thirdly, to make this method compatible with the limitation of the NISQ hardware, we formulate a time-dependent variational simulation of the QITC. Hence, one derives a new set of equations for time-dependent updates of parameters for an ansatz circuit. Finally, we test various systems like the diatomic molecules, 2D Heisenberg model, Sherrington–Kirkpatrick model, and a spin model constructed from a 3-SAT annealing problem to ensure the existence of the speedup is general among many different cases. In addition, through numerical examples, we also demonstrate that the newly proposed variational version of the QITC not only converges faster than the standard variational QITE but also manifests higher robustness against noises of moderate strength. With a properly chosen set of control Hamiltonians, the variational simulation of the QITC does not incur many measurement overheads. For instance, if one chooses $H_d$ that commutes with the constituting Pauli terms in $H_p$ then one can significantly minimize the number of measurement overheads.

\section*{Results}
\subsection*{Theoretical Motivation} \label{appgeneralund}
A time-dependent Hamiltonian in the imaginary-time domain is given by,
\begin{equation}\label{ITE}
\frac{d\psi(\tau)}{d\tau} = -(H_p-E_\tau)\psi(\tau),
\end{equation}
where $\psi(\tau)$ is the system's state vector, $H_p$ is the problem Hamiltonian, and $E_\tau=\expval{H_p}{\psi(\tau)}$ is the state's expected energy, and it is introduced in Eq.~\ref{ITE} to ensure the time-evolved wave function will be properly normalized.
By eigendecomposion, we can write the problem Hamiltonian $H_p$ as
$H_p=U \Lambda U^\dagger$, where $U$ is a unitary matrix and $\Lambda=diag(\lambda_0,\cdots,\lambda_n)$ is a diagonal matrix where $\lambda_0\leq \lambda_1 \leq \cdots \leq \lambda_n$. Consider an additional Hamiltonian $H_d$ that commutes with $H_p$ and follows the same ordering of eigenstate in energy then,
\begin{equation}
H_d =U D U^\dagger,
\end{equation}
where $D=diag(e_0, \cdots ,e_n)$ is a diagonal matrix with $e_0 \leq e_1\leq \cdots \leq e_n$. The energy difference between the ground state and the i-th eigenstate will change from $\Delta_{i0}=\lambda_i-\lambda_0$ to $\tilde{\Delta}_{i0}=\Delta_{i0}+(e_i-e_0)$. Since the extra contribution to the energy gap $\delta_{i0}=e_i-e_0>0$ for all $\delta_{i0}$, the imaginary time evolution under $H_p+H_d$ will then be exponentially accelerated (proportional to $\exp(\delta_{i0})$) in comparison to the time evolution under $H_p$ alone. Through simple arguments, we illustrate a sufficient (but not necessary) condition that may yield the desired exponential speedup on the imaginary-time evolution by imposing an additional Hamiltonian $H_d$ that commutes with $H_p$ and preserve the ordering of eigenstates of $H_p$ by energy. Even though it is computationally prohibitive to construct $H_d$ when the Hilbert space is large \cite{Seki2021}, the revelation in this theoretical analysis provides a clear direction to proceed. We carefully discuss a series of practical approximations attempting to gain speedup in the imaginary-time evolution. For all the simulation experiments in the article, the construction of control Hamiltonian follows the approximation methods. We also denote that following the control strategy involving maximum admissible control impulses assist to ensure the substantial acceleration effect. 

\begin{figure*}[htb]
\begin{center}
\includegraphics[width=1\textwidth]{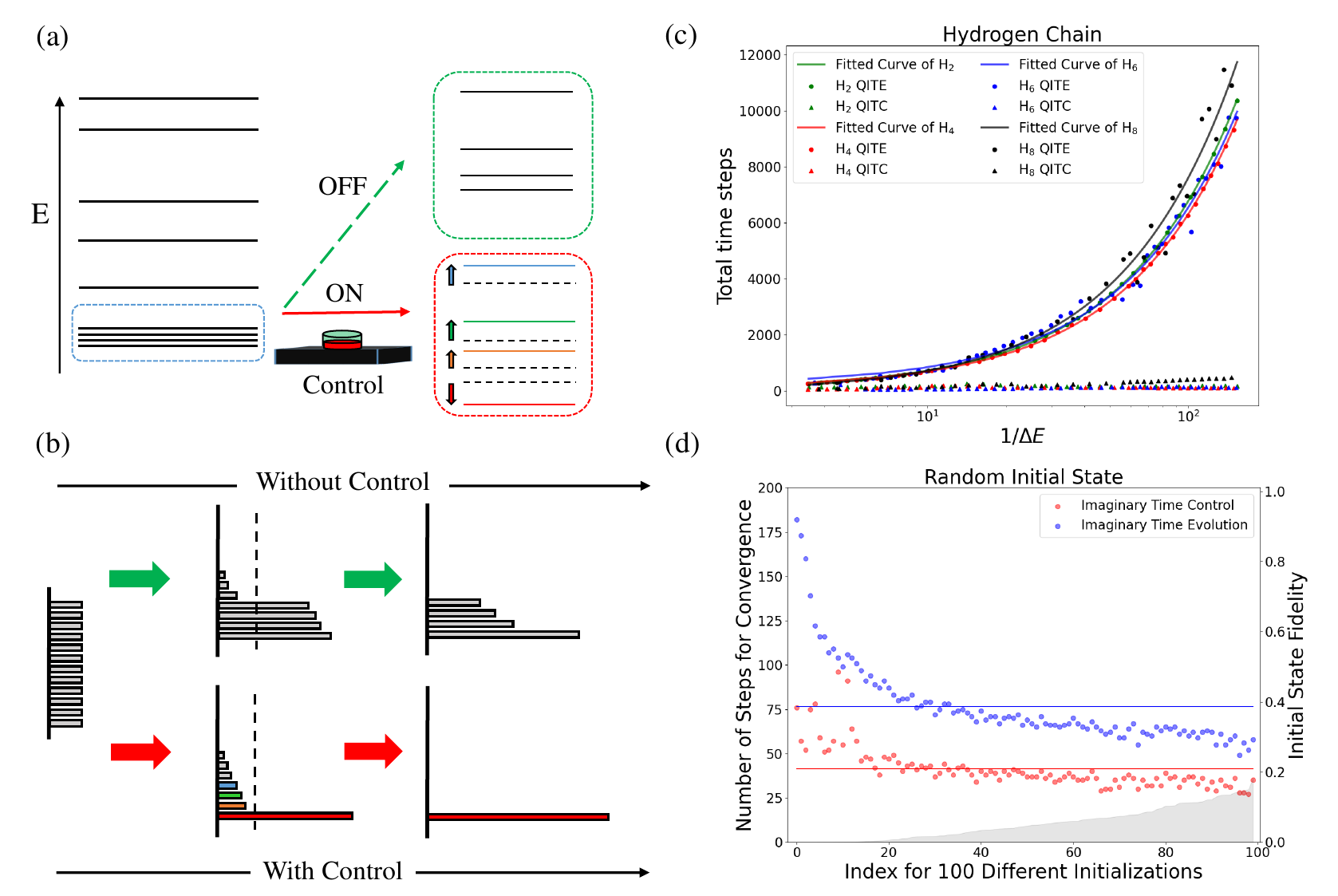}
\caption{(a) A temporally modulated Hamiltonian to steer many-body quantum dynamics (b) The QITE gets slowed down significantly when the current state is the combination of many low-energy eigenstates. (c) The H-chain results for even Hydrogens that are scaled from $H_2$ to $H_8$. The dots of different colors indicate the total convergence steps of QITE and QITC in different systems, and lines of different colors indicate that the fitted curves of QITE satisfy $f(x)=ae^x+b$, where $x=log(1/\Delta E)$. On the other hand, the QITC results yielded $f(x)=ax+b$. (See the details in Supplementary Section B1) (d) The numerical results of 4 qubits $H_2$ molecule with bond length 0.74 and 100 random initial state, the x-axis is the case number we assign for randomly chosen initial points, the left y-axis is the number of convergence, and the right y-axis is the state fidelity between the initial state and ground state. The control Hamiltonian used here is single Pauli $Z$.} \label{2}   
\end{center}
\end{figure*}
\subsection*{Quantum imaginary time control}
A quantum system's dynamical evolution under a time-dependent Hamiltonian in the imaginary-time domain is given by the modified time-dependent Schrodinger's equation,
\begin{equation}\label{H}
\frac{d\psi(\tau)}{d\tau} = -(H_p+\beta(\tau)H_d-E_\tau)\psi(\tau),
\end{equation}
where $\psi(\tau)$ is the system's state vector, $H_p$ is the problem Hamiltonian, and $H_d$ is the control Hamiltonian coupled to a time-dependent control pulse $\beta(\tau)$. $E_\tau=\expval{(H_p+\beta(\tau)H_d)}{\psi(\tau)}$ is the state's expected energy, and it is introduced in Eq.~\ref{H} to ensure the time-evolved wave function will be properly normalized.
According to the Lyapunov method and La Salle invariance principle \cite{Alessandro2008}, the state preparation problem can be re-formulated as an optimization problem in which the target state minimizes a Lyapunov function $V(\psi(\tau))$, which must satisfy the following conditions. Consider a system of differential equations $\dot{\psi}(\tau)=f(\psi(\tau))$ with a smooth $f$ and the state of the system satisfies the conservation of probability $||\psi(\tau)||=1, \forall \tau \geq 0$, which means that $\psi$ is on the unit sphere $\mathbf{S}=\{x\in\mathbf{C}^n:||x||=1\}$. Consider a smooth function $V(\psi)$ on the phase space $\Omega$, such that $V(\psi)\geq0$ and $\frac{d V(\psi)}{d\tau}\leq 0$ for $\psi \in \Omega$. Let us define $\mathcal M$ to be the set of points $\psi \in \Omega$ such that $d V(\psi)/d\tau=0$, then every solution of the time-dependent Schrodinger's equation converges to $\mathcal{M}$ as $t\rightarrow \infty$.

For the preparation of the ground state of a Hamiltonian, the Lyapunov function can be chosen as follows,
\begin{equation}\label{V}
V(\psi(\tau))=\expval{P}{\psi(\tau)},
\end{equation}
where $P=H_p-\tilde{E}$ with $\tilde E$ a constant shift of energy to ensure that $P$ is a semi-positive definite Hermitian operator\cite{Grivopoulos2003}. The time derivative of the Lyapunov function in Eq.(\ref{V}) reads,
\begin{align}\label{dV}
\dot{V}(\psi)&=2\sigma_{H_p}^2(\tau)-\beta(\tau)D(\tau),
\end{align}
where $\psi$ is an abbreviation for $\psi(\tau)$. The other variables are given by
\begin{equation}
\begin{array}{l}\sigma_{H_p}^2(\tau) \equiv\left\langle\psi\left|H_{p}\right| \psi\right\rangle^{2}-\left\langle\psi\left|H_{p}^{2}\right| \psi\right\rangle,
\\ T(\tau) \equiv\left\langle\psi\left|\left\{H_{p}, H_{d}\right\}\right| \psi\right\rangle-2\left\langle\psi\left|H_{p}\right| \psi\right\rangle\left\langle\psi\left|H_{d}\right| \psi\right\rangle,\end{array}
\end{equation}
where $\{\cdot,\cdot\}$ is the anti-commutator. In this case, the Lyapunov-controlled quantum dynamics will be driven to the asymptotically stable points, which form the ground-state manifold, in $\mathcal{M}$. To make $\dot{V}(\psi) \leq 0$, it is sufficient to enforce $\beta(\tau)T(\tau)\geq 0$ as $\sigma_{H_p}^2(\tau)$ is less than or equal to zero. In summary, a successful imaginary-time control to prepare the ground state of a Hamiltonian is to design an appropriate $\beta(\tau)$ and suitable $H_d$ (see Supplementary Section A). For this work, we proposed some control strategies (see Method) that are inspired by the bang-bang and approximate bang-bang Lyapunov control that can provide rapid state transitions for quantum systems in the real-time domain\cite{Kuang2014}.
\begin{figure}[!htb]
\includegraphics[width=0.48\textwidth]{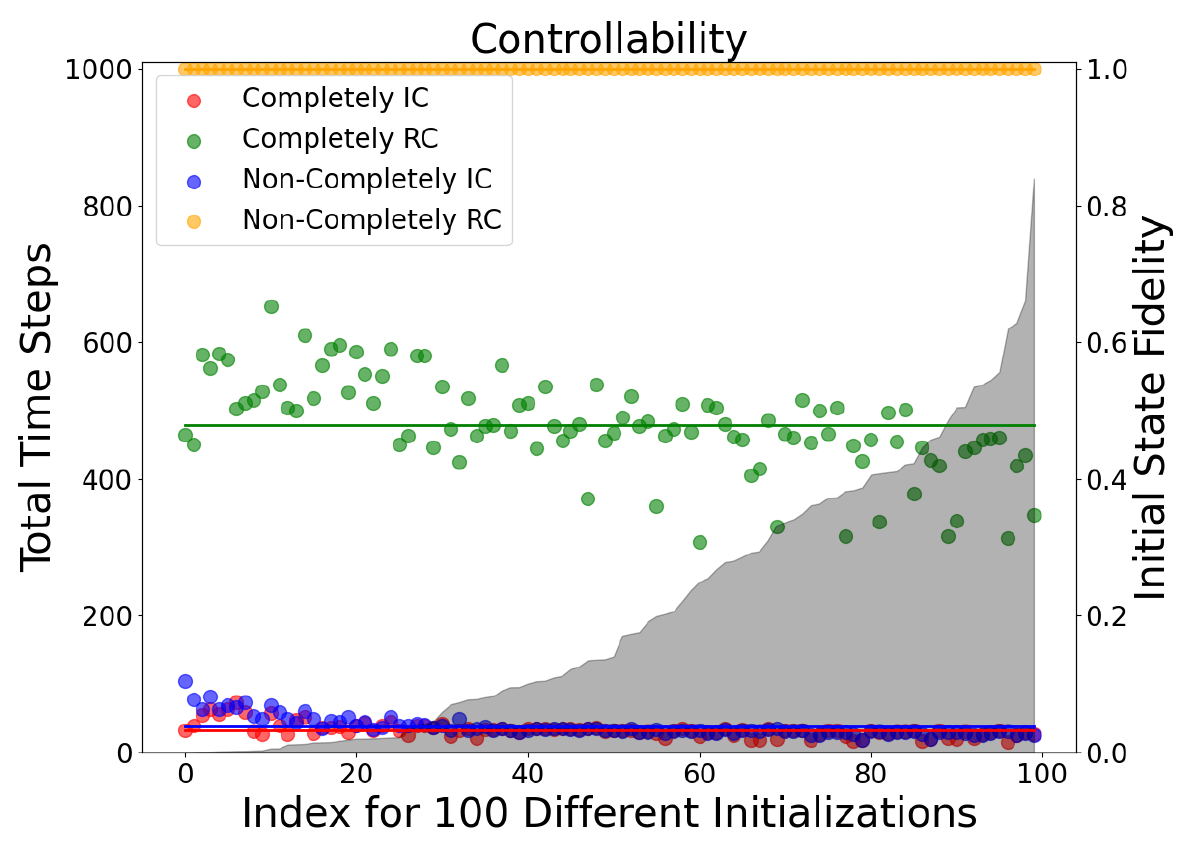}
\caption{The numerical result of 2 qubits $H_2$ molecule (see Supplementary Section D), the x-axis is a serial number of initial states, the left y-axis is the convergent steps, the maximum number of steps is limited to 1000, and the right y-axis(corresponding to the grey area) is the initial and ground state overlap. The complete imaginary time control(Completely IC (red)), the non-completely imaginary time control(Non-Completely IC (blue)), and the complete real-time control(Completely RC (green)) could all converge in time. However, the non-completely real-time control(Non-Completely RC (yellow)) cannot converge in a limited time.
} \label{1}   
\end{figure}
\subsection*{Comparison to imaginary time evolution}
We find that QITE gets slowed down significantly when the current state is mainly a superposition of many densely spaced low-energy eigenstates Fig.~\ref{2}(b).
For the original $H_p$ the energy gap between the ground state and the first excited state remains fixed and limits the simulation efficiency. Consistent with the theoretical motivation stated before, we find that the QITC temporally modulates the entire spectrum of $H(\tau)$ during the evolution and finally returns back to the original spectrum of $H_p$. Those temporally enlarged energy gaps contribute to the acceleration of the simulation process, see Fig.~\ref{2}(a)

In Fig.~\ref{2}(c), we present the results of ground-state preparation, it shows that the QITC provides orders of magnitude speedup (suggesting exponential-like acceleration) in the rate of convergence with respect to the intra-molecular distance. Note that the $\Delta_{10}$ energy gap reduces when the intra-molecular distance grows. This observed scaling trend certainly benefits the simulation of large complex systems with small energy gaps. To ensure this observed significant speedup holds in a variety of chemical systems, we provide simulation results of four, eight, twelve, and sixteen-qubit Hydrogen chains, a twelve-qubit LiH, and HF in the following Section.
In Fig.~\ref{2}(d), we illustrate the enhanced convergence efficiency for the QITC with different initial states. These results also indicate that the convergence of the QITC does not sensitively depend on the initial states. 

\subsection*{Comparison to the real-time control for the ground-state preparation}
An essential question for a driven state preparation concerns the controllability for the given control Hamiltonians $\{H_d^0,H_d^1,...\}$, i.e. whether the quantum system can be driven to the ground state of $H_p$ from any given initial state when it is subjected to evolve under the time-dependent Hamiltonian $H(\tau)=H_p + \beta(\tau)H_d$ in our context,  where $H_d = \sum_{i}\beta^i(\tau)H_d^i$. Without loss of generality, we write $\beta(\tau)H_d$ as opposed to the more general form $\sum_{i}\beta^i(\tau)H_d^i$ throughout this work. This simple question for real-time control turns out to be rather difficult to answer for a large quantum system. A common technique involves analysis of the structure and rank of corresponding Lie groups and algebra for the propagators\cite{Schirmer2001, Albertini2003, Huang1983}. It is computationally demanding to determine the controllability of a particular setup, and it is unlikely that a random selection of control Hamiltonians can guarantee complete controllability. The challenge to select an appropriate set of control Hamiltonians (with respect to a given initial state) poses a severe challenge to derive a practical real-time control strategy to prepare a target ground state.

The same question regarding controllability admits a much clearer answer in the imaginary-time domain. As long as the imaginary-time-evolved state $\psi(\tau)$ and the ground state have a non-zero overlap, the system can always converge to the ground state in a sufficiently long evolution time. 
Hence, a more crucial question is whether a set of control Hamiltonians along with the corresponding control strategy $\beta(\tau)$ (i.e. to ensure $\dot{V}(\psi)\leq 0$) can substantially accelerate the driving from a given initial state to the ground state of $H_p$. As discussed next, the imaginary-time Lyapunov control can indeed work with a broader range of control Hamiltonians for accelerating the ground-state preparations.

We further illustrate the differences between real-time and imaginary-time Lyapunov control with a numerical example involving $H_2$ molecule in Fig.~\ref{1}. We randomly choose 100 initial states and examine how long the real-time and imaginary-time evolutions converge to the ground state under control Hamiltonians with various degrees of controllability. When the control Hamiltonian satisfies the strongly complete controllability\cite{Zhang2005}, we expect that the driven dynamics converge to the ground state without difficulty. While this expectation holds in the numerical experiments, we find the imaginary-time evolution converges much faster (with roughly one-tenth of the time steps for the real-time evolution on average). We expect this performance gap to be further enlarged with the system size.  For control Hamiltonians with incomplete controllability, the imaginary-time control can still lead to satisfactory convergence in a small number of time steps. In contrast, the real-time simulation under the same control Hamiltonians cannot converge for all 100 initial states in the pre-defined maximum number of steps allowed.
\subsection*{Comparing QITC and VQE in Noisy Environment}
To make the method accessible in the noisy intermediate-scale quantum era, we further propose a variational form of the algorithm that could work with shallow quantum circuits. 
\begin{figure}
\centering
\includegraphics[width=0.48\textwidth]{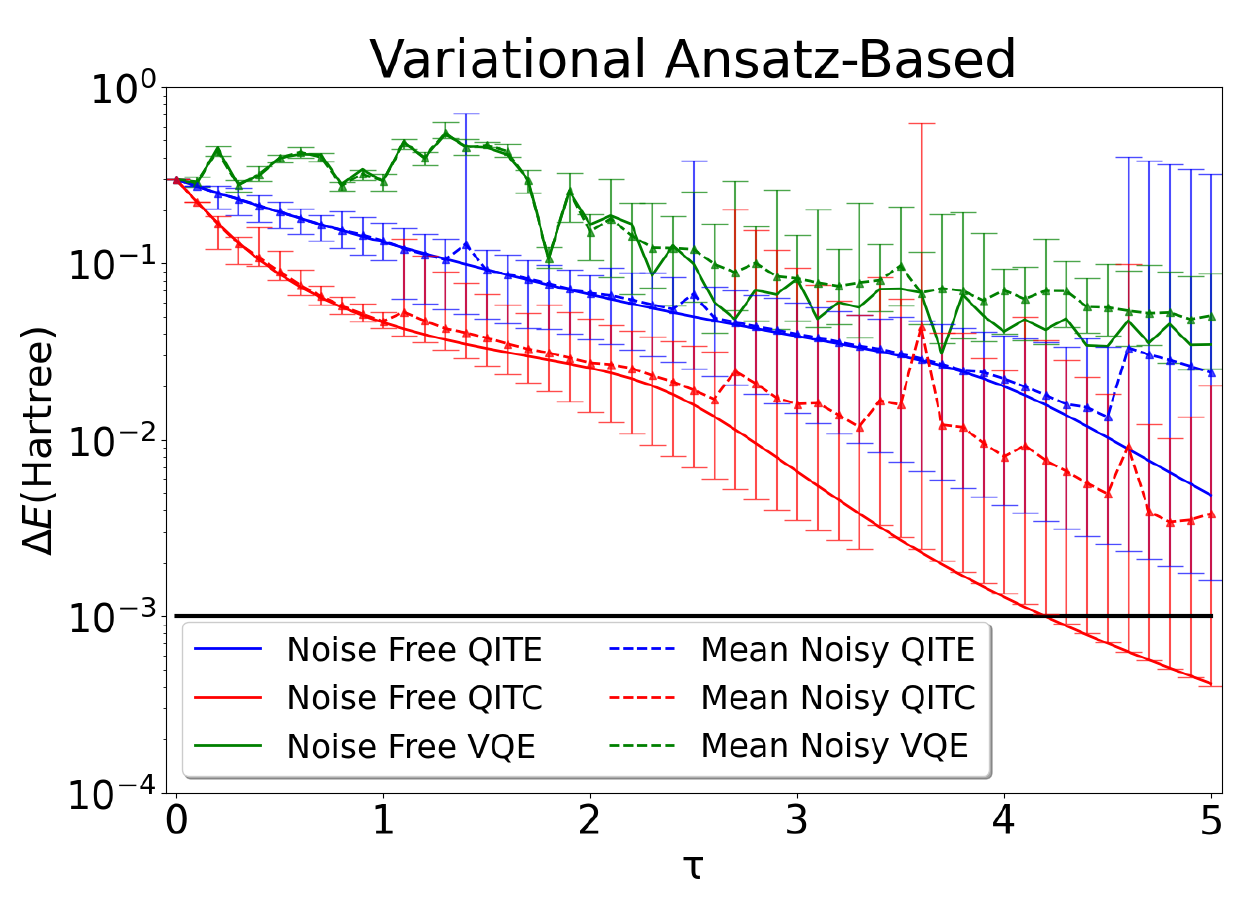}
\caption{The numerical results of 4-qubit Hydrogen molecule (see Supplementary Section D) with a bond length of 0.74$\r{A}$. The x-axis is imaginary time with a single time-step $\Delta\tau=0.1$ and the total time step is 50. The y-axis is the energy difference from ground-state energy. The solid line is the result of the numerical noise-free model (the black line is chemical accuracy). The numerical noisy model is simulated by adding the single qubit depolarization error channels with parameter $\lambda=10^{-4}$ and the CNOT depolarization error channels with parameter $\lambda=10^{-5}$ (We use TensorCircuit\cite{Zhang2022} to effectively get this result.). The dashed line is the mean value of 50 noisy results, and the error bar is bounded by the worst and best simulation results. The control strategy used in this simulation is approximately bang-bang control with $S=0.3$ and $\gamma=5$.
} \label{3}   
\end{figure}
In this section, we compare the simulation ability of VQE, variational imaginary time evolution, and variational imaginary time control. In Fig.~\ref{3}, we compare the results among the VQE, the variational ansatz-based QITE, and the variational ansatz-based QITC for the ground-state preparation of a 4-qubit $H_2$ system under both noise-free and noisy situations. We adopt the fully connected ansatz with high expressibility\cite{Sim2019} initialized by the same random initial parameters for all the methods. The results show that the QITC converges faster than the QITE and the VQE for both the noisy and noise-free models. The control Hamiltonian we using here is the single Z and double Z selected from the hydrogen Hamiltonian $H_p$.
\begin{table}[!htb]\scalebox{0.8}{
\begin{tabular}{|l|l|l|l|l|}
\hline
Model                & $(1\times1)$& $(2\times2)$& $(3\times3)$& $(2\times2)_{D}$ \\ \hline
h/J                  & 0.1/0.09 & 0.1/0.09  & 0.1/0.09  & 0.2/0.1    \\ \hline
QITE total time step & 366      & 764       & 5963      & 28934      \\ \hline
QITC total time step & 130      & 253       & 1137      & 182        \\ \hline
Difference           & 236      & 511       & 4826      & 28752      \\ \hline
Ratio                & 2.82     & 3.02      & 5.24      & 158.98     \\ \hline
Control              & ZIIZ     & ZIIZ      & ZIIZ      & ZIIZ       \\
Hamiltonian ($H_d$)  &          & ZIZ       & ZZZ       & ZZZ         \\ \hline
Number of $H_d$      & 4       & 18       & 32       & 18  \\ \hline
\end{tabular}}
\caption{The result of 2D Heisenberg model. The $(2\times2)_{D}$ is the 'difficult' regime in the space of Hamiltonian parameters case where QITE required a great number of steps to converge. All the control Hamiltonians are in the structure of simple Pauli $Z$ with the cyclic structure mentioned above and we only write down the non-trivial part of Pauli in the table. For example, the ZIZ in the 2×2 model is the abbreviation of ZIZIIIIII. The number of control steps for all cases is 100 steps.
}\label{T1}
\end{table}

\subsection*{Examples Beyond H Chains}
\begin{figure*}[!htb]
\begin{center}
\hspace*{-0cm}\includegraphics[width=1\textwidth]{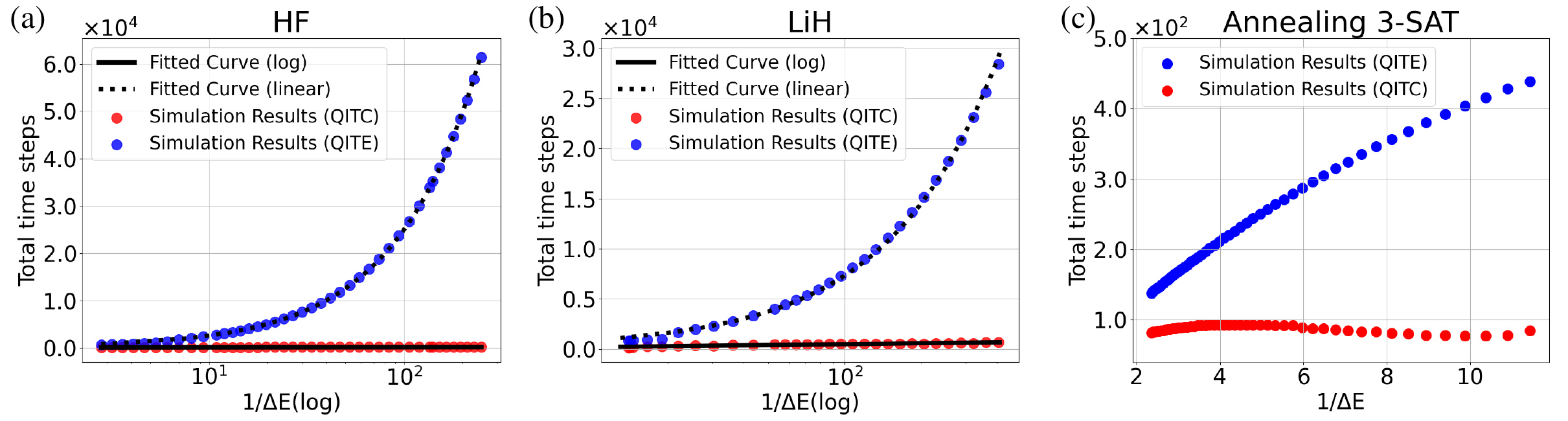}
\caption{(a)The numerical result of 12 qubits HF molecule (see Supplementary Section D) with bond length from 1.00 to 2.36 ${\AA}$ and the same initial state with equal superposition of all the basis states, the QITE is the result of imaginary time evolution. The QITC is the result of the imaginary time control, the x-axis refers to the log of one over the energy gap between the ground state and the first excited state, and the y-axis is the total time step. QITE fitted curve f(x)=248.67 exp(x) + 126.12 and QITC fitted curve f(x)=12.80 x+ 110.09, where $x=log(1/\Delta E)$. (b) The result of LiH problem. The x-axis is log of one over energy gap between the ground state and the first excited state. The y-axis is the total time steps of the ITE and ITC. QITE fitted function f(x)=71.60 exp(x) + 154.96 and QITC fitted curve f(x)=125.31 x – 108.81, where $x=log(1/\Delta E)$. (c)The result of 3-SAT quantum annealing problem. The x-axis is the one over the energy gap between the ground state and the first excited state. The y-axis is the converge steps difference between the ITE and ITC.
}\label{diatomic}
\end{center}
\end{figure*}
First, we simulate two diatomic molecules HF and LiH to test the speedup from the control in molecule systems. In Fig. \ref{diatomic} (a-b), we show that both the HF and LiH using the same control Hamiltonian used in the H-chain system can obtain exponential-like speedup. 
Second, we consider a spin model constructed from the annealing solving of an 11-qubit 3-SAT problem during the linear schedule, see $H(s)$ defined in Supplementary Section B3. We then implement and compare the simulations of the ITC and the ITE in order to prepare the ground states for a series of $H(s)$ chosen along the adiabatic path. According to the simulation result in Fig. \ref{diatomic} (c), we numerically verify the speed-up of the ITC as the instantaneous energy gap $\Delta E(s)$ shrinks.
Third, we demonstrate the 2D Heisenberg model  (see Hamiltonian details in the Method Section)  with different system sizes using control Hamiltonian constructed only by Pauli $Z$ up to cube order. As we can see from the results in Table \ref{T1}, Lypapunov control provides obvious speed-up while the system size scales up. The $(2\times2)_{D}$ case also indicates the existence of great speed-up using simple control Hamiltonian Pauli $Z$.

\begin{figure}[ht]
\begin{center}
\includegraphics[width=0.48\textwidth]{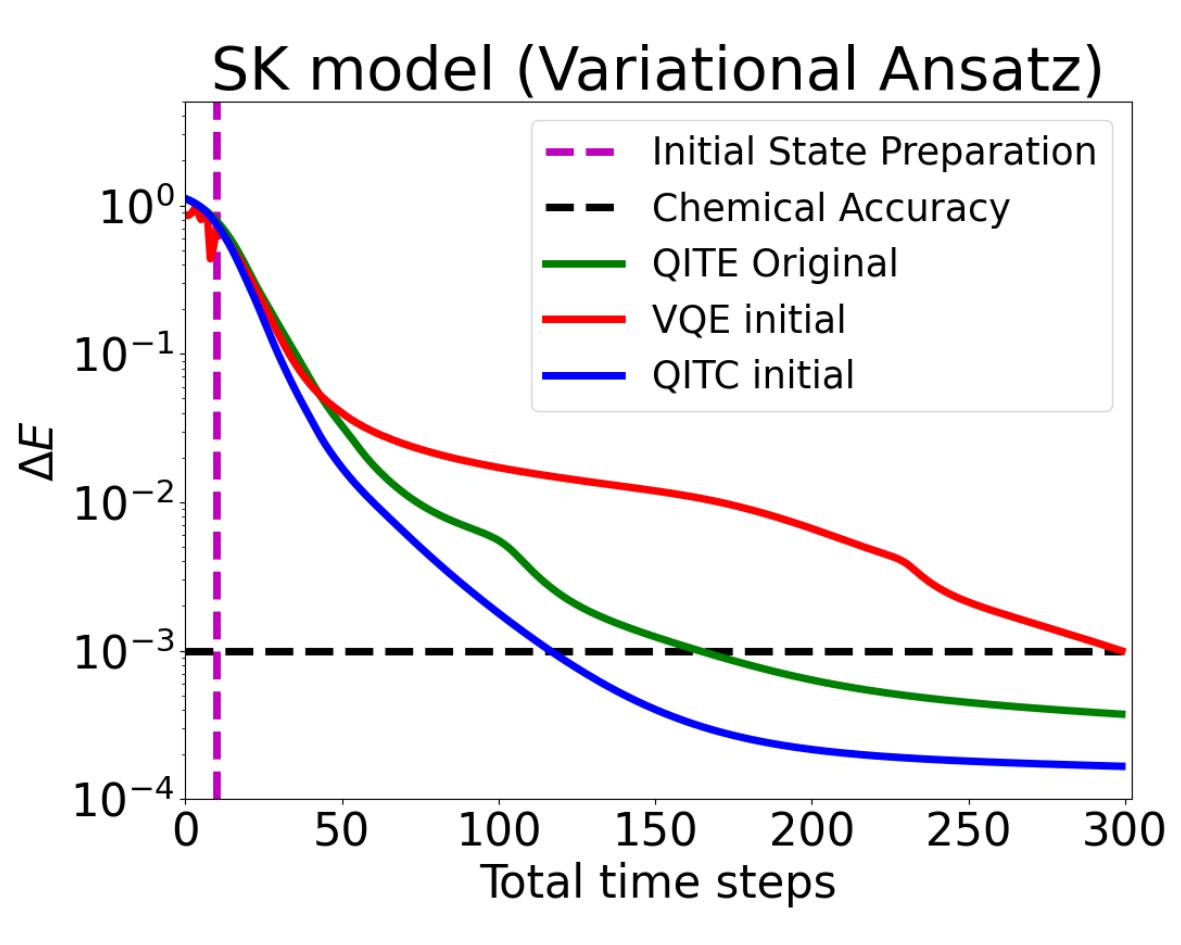}
\caption{The result of 4 qubits SK model. The x-axis is the number of time steps and the y-axis is the energy difference from the ground state. The control initial can achieve chemical accuracy around three times faster than the VQE initial with a smoother convergence path.}\label{SKVQE}
\end{center}
\end{figure}

Finally, we simulate a 4-qubit variational-ansatz based imaginary time evolution of the spin glass model (see Hamiltonian details in the Method Section) in Fig. \ref{SKVQE} with random variables $J_{ij}=(0.049, 0.215, 0.103, 0.045, -0.076, 0.146)$, the energy difference between the ground state and first excited state $\Delta_{10}$, in this case, is 0.22 which is not small compared to the result of our molecule models. Although the QITE can work fine with such $\Delta_{10}$, it is still difficult to find a good initial state for a complex system like this SK model. To prepare a better initial state, we use VQE with COBYLA optimizer and variational QITC with commute basis $H_d=\{XXXX, YYYY, ZZZZ\}$ at the first ten steps respectively. After the initial state preparation, we use variational imaginary time evolution to evolve to the ground state given these two initial states. It can be seen that compared with the VQE initial state, the control initial state can achieve higher accuracy under the same number of convergence steps. The details and more discussion of the models are presented in Supplementary Section B.

\subsection*{Resource Consumption for scaling H-chain system} \label{resource}
\begin{figure}[!htb]
\begin{center}
\hspace*{-0cm}\includegraphics[width=0.5\textwidth]{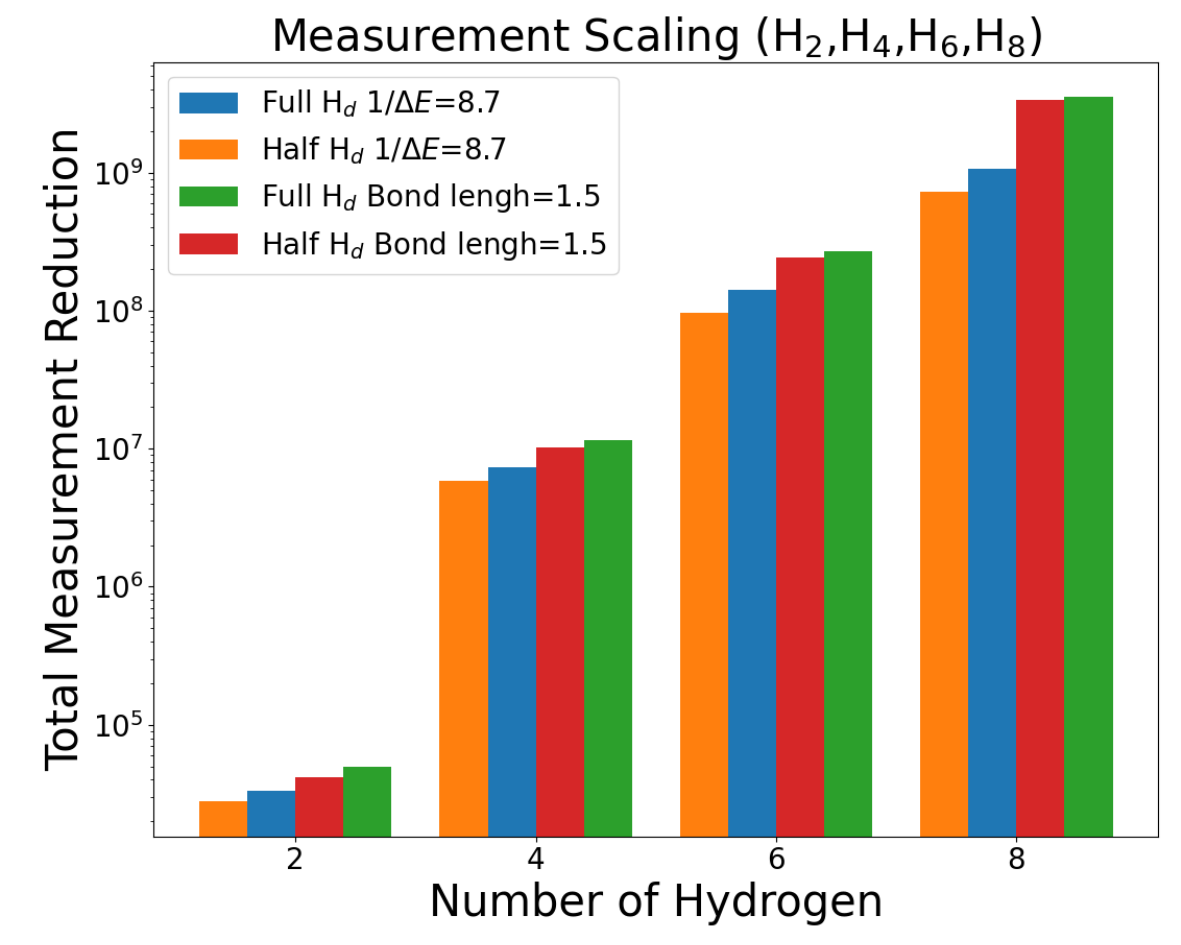}
\caption{Total measurement difference results. The results are the estimation of the difference in total measurements for imaginary time evolution and imaginary time control convergence, the calculation details are listed in Table \ref{T2}.  The x-axis is the number of H in the system and the y-axis is the log plot of the measurement reduction. The Full $H_d$ and Half $H_d$ are listed in Supplementary Section B.
}\label{scalingtest}
\end{center}
\end{figure}

So far, we have only discussed the conceptual advantage of our method (i.e. faster convergence towards the ground state). As we propose this method in the context of digital quantum simulation on a quantum computer, we further analyze how our method helps to reduce the consumption of quantum resources.
Here we present the number of measurements for $H_2, H_4, H_6$ and $H_8$ (i.e. 4, 8, 12, and 16 qubits)). In Fig. \ref{scalingtest}, we summarize this reduction in the number of measurements with different choices of $H_d$(see Supplementary Section B for details): The Full $H_d$ with the same $1/\Delta E$(blue bar), the Half $H_d$ with same $1/\Delta E$(orange bar), the Full $H_d$ with the same bond length(green bar) and the Half $H_d$ with the same bond length(red bar). The result shows that all of them give some polynomial reductions in the number of measurements, compared to the standard variational ansatz-based imaginary time evolution. This is because the extra measurement numbers from ${H_d, H_p}$ are relatively small in comparison to the variational ansatz update measurements, which scales as $N_pN_\theta$, where $N_p$ is the number of Pauli terms of $H_p$ and $N_\theta$ is the number of the parameter in the ansatz circuit (we used k-UpCCGSD from PennyLane for the reference). In Table 
\ref{T2} we show the $N_p$, $N_\theta$, and extra measurements for different $H_d$. Thus, according to Fig. \ref{scalingtest}, for the molecules tested in this study, we find the measurement resource reduction increases polynomially with the system size. 
\begin{table*}[]
\centering
\begin{tabular}{|c|c|c|c|cc|cc|cc|}
\hline
\multirow{2}{*}{} &
  \multirow{2}{*}{$N_p$} &
  \multirow{2}{*}{$N_\theta$ k-UpCCGSD} &
  \multirow{2}{*}{$N_p N_\theta$} &
  \multicolumn{2}{c|}{Full Hd} &
  \multicolumn{2}{c|}{Half Hd} &
  \multicolumn{2}{c|}{Same Bond Length} \\ \cline{5-10} 
 &
   &
   &
   &
  \multicolumn{1}{c|}{$\{H_d,H_p\}$} &
  $\Delta Step$ &
  \multicolumn{1}{c|}{$\{H_d,H_p\}$} &
  $\Delta Step$ &
  \multicolumn{1}{c|}{$\Delta Step_f$} &
  $\Delta Step_h$ \\ \hline
$H_2$ & 15   & 6 (k=1)   & 90      & \multicolumn{1}{c|}{5}      & 381 & \multicolumn{1}{c|}{5}     & 327 & \multicolumn{1}{c|}{554}  & 473  \\ \hline
$H_4$ & 185  & 72 (k=2)  & 13320   & \multicolumn{1}{c|}{1006}   & 555 & \multicolumn{1}{c|}{577}   & 449 & \multicolumn{1}{c|}{872}  & 771  \\ \hline
$H_6$ & 919  & 270 (k=3) & 248130  & \multicolumn{1}{c|}{16627}  & 572 & \multicolumn{1}{c|}{8804}  & 399 & \multicolumn{1}{c|}{1098} & 991  \\ \hline
$H_8$ & 2913 & 672 (k=4) & 1957536 & \multicolumn{1}{c|}{112396} & 544 & \multicolumn{1}{c|}{57966} & 377 & \multicolumn{1}{c|}{1804} & 1712 \\ \hline
\end{tabular}
\caption{Summary of Measurement Test, where $N_p$ is the number of Pauli terms of problem Hamiltonian $H_p$ (here we use Jordan-Wigner transformation), $N_\theta$ is the number of parameters in the ansatz circuit (here we use k-UpCCGSD) and $N_pN_\theta$ is the number of measurement to update the variational-ansatz based imaginary time evolution. $\{H_p, H_d\}$ is the extra measurement for update control Hamiltonian coefficient $\beta(\tau)$ and $\Delta Step$ are the total time step difference between imaginary time evolution and imaginary time control. The $\Delta Step_f$ and $\Delta Step_h$ are the total time step difference using Full $H_d$ control and Half $H_d$ control for same bond length case.}\label{T2}
\end{table*}

\section*{Discussion}
In summary, we propose to utilize the imaginary-time Lyapunov control to prepare ground states and explain the advantages of QITC. First, imaginary time control can speed up imaginary time evolution with the proper design of control Hamiltonian and control function $\beta(\tau)$. Through numerical experiments on a broad spectrum of realistic models, we show that compared to standard imaginary-time evolution, imaginary-time control provides substantial speed-up for all systems. And for the selected small-gap systems invested in this work, an exponential-like speedup is observed. Secondly, compared to real-time control, imaginary-time control admits more relaxed conditions for controllability (when the target is the ground state) and a broader range of control Hamiltonians to facilitate the desired state transitions in a finite time. Thirdly, to make the present method accessible in the NISQ era, we propose a variational simulation of the QITC with an ansatz circuit. Finally, we show various examples to strengthen the speedup from QITC. We also show that when the control Hamiltonian is chosen appropriately, it does not incur many additional measurement costs and exhibits higher robustness against noises. These merits make the present approach a natural replacement for the variational ground-state preparation. For future work, we aim to study imaginary time control for other challenging state preparation tasks, such as excited states simulation and Gibbs state preparation, etc. 

\section*{Methods}
In this section, we present the details of the variational algorithm for QITE simulations, studied spin-model Hamiltonian, and the imaginary time control strategies.

\subsection*{Variational algorithm for QITE simulations}
To utilize the proposed imaginary-time control to prepare a ground state on a near-term quantum device, we rely on the time-dependent variational principle to approximate the evolution of the QITC as sequential updates of parameters for an ansatz circuit, as first proposed by McArdle et al.,\cite{McArdle2019} for the QITE. The modified algorithm proceeds as follows. Given a time-dependent Hamiltonian $H(\tau)=H_p+\beta(\tau)H_d$, we would invoke the
the McLachlan's variational principle\cite{McLachlan1964,Broeckhove1988},
\begin{equation}
\delta||(\partial/\partial\tau+H(\tau)-E_\tau)\ket{\psi(\tau)}||=0,
\end{equation}
which conducting the dynamic evolution $\partial \left|\psi(\tau)\right>/\partial \tau =-\left[H(\tau)-E_\tau\right]\left|\psi(\tau)\right> $
by deducing the incremental update (corresponding to one time step $\delta \tau$) of the parameters $\theta$ for an ansatz circuit. Following Ref.~\cite{McArdle2019}, we need to solve the following equations,
\begin{equation}
\sum_j A_{ij}\dot{\theta_j}=C_i, \forall i
\end{equation}
where,
\begin{equation}
\begin{array}{c}A_{i j}=\Re\left(\frac{\partial\langle\phi(\tau)|}{\partial \theta_{i}} \frac{\partial|\phi(\tau)\rangle}{\partial \theta_{j}}\right) \\ C_{i}=\Re\left(-\sum_{\alpha} \lambda_{\alpha} \frac{\partial\langle\phi(\tau)|}{\partial \theta_{i}} h_{\alpha}|\phi(\tau)\rangle\right.\end{array}
\end{equation}
and $h_\alpha$ and $\lambda_\alpha$ are the Pauli terms and coefficients of the Hamiltonian $H=H_p+\beta(\tau) H_d=\sum_\alpha \lambda_\alpha(\tau) h_\alpha$. With the new ansatz state, we can then evaluate the Lyapunov function and assign values to the pulse $\beta(\tau)$ via a pre-determined control rule to keep $\dot{V}(\psi)<0$. The procedure of alternating updates of $\theta$ and $\beta$ is then repeated until a fixed point is reached.

\subsection*{Hamiltonian details for the spin models}
The 2D Heisenberg models are constructed with non-periodic boundary conditions, whose Hamiltonian can be written as 
$$H_p=h_i\sum_i Z^i+J_{ij} \sum_{edge}X^iX^j+Y^iY^j+Z^iZ^j.$$
The model information in the table presents edge $\times$ edge, for example, $2\times 2$ means $3\times 3$ lattice.
The Hamiltonian of Sherrington–Kirkpatrick model we use can be written as  $$H_p=\sum_{i<j}J_{ij}(X^iX^j+Y^iY^j+Z^iZ^j)$$,
with random generate variables $J_{ij}\in(-0.5,0.5)$ and using $\ket{++\dots+}$ as the initial state. 
\subsection*{The imaginary time Control Strategy} \label{appsatbbistrategy}
In this section, we will discuss the approximate bang-bang control and bang-bang control laws we use to design the $\beta(\tau)$ for imaginary time control and the inverse control strategy we used to accelerate our molecule system ground state convergence. To appreciate our approach, we first review some typical real-time control strategies designed to guarantee $\dot{V}\leq 0$. Following a standard convention \cite{Kuang2014}, we refer to the following choice as the standard Lyapunov control,
$$
\beta_k(t)=-K_kT_k(t),
$$
where $\beta_k(t)$ is an external real-valued control field, $K>0$ is the control gain used to adjust the amplitude of the control field, and $T_k(t)\equiv (\expval{i[H_d, H_p]}{\psi}$ for the real-time control. Another commonly used strategy is the bang-bang Lyapunov control,
$$\beta_k(t)=\left\{
\begin{array}{rcl}
-S,& & (T_k>0)\\
 S,& & (T_k<0)\quad k=1,...,m,\\
 0,& & (T_k=0)
\end{array} \right.
$$
where $S>0$ is the maximum strength of the control field. In order to achieve a good trade-off between convergence and the rapidity of control, Kuang et al.\cite{Kuang2014} propose an approximate bang-bang control as follows,
$$
\beta_k(t)=\frac{2S}{1+e^{-\gamma T_k}}-S.
$$
where $\gamma>0$ is a parameter used to adjust the hardness of the control strategy.
For the imaginary time control strategy, let us generalize the standard Lyapunov control such that it could work with the imaginary-time evolution. We redefine $T_k\equiv 2\expval{H_p}{\psi}\expval{H_d}{\psi}-\expval{\{H_p,H_d\}}{\psi}$ in this case. The $H_d$ related terms in $T_k(\tau)$ may entail lots of extra measurements if they cannot be obtained by measuring the Pauli terms appearing in $\expval{H_p}{\psi}$. To reduce the measurement cost and still maintain a powerful $H_d$ to provide an enhanced convergence, we propose the following strategy. We first decide if the state in the quantum circuit has high overlap with any eigenstate of $H_p$ or $H_d$ by checking the value of $T_k$, if $T_k < L$ then we do not apply any control pulses, otherwise we use a similar control strategy for the real-time case introduced above.
$$\beta_k(\tau)=\left\{
\begin{array}{rcl}
&\frac{2S}{1+e^{-\gamma T_k}}-S,&  (T_k\geq L)\\
&0,&  else
\end{array} \right.
$$
where L is some pre-defined threshold value. If the state is close to an eigenstate (i.e. $T_k(\tau)<L$), we should turn off the control field and let the system evolves under $H_p$ in the imaginary-time domain. This truncation can greatly reduce the measurement costs (for the implementation of the corresponding variational algorithm) in the region where the state will linearly converge to the eigenstate. Finally, we test how the truncation (i.e. setting $\beta_k(\tau)=0$ when $T_k(\tau)<L$) will affect the precision of the converged results given by truncating the control pulse, and we also test the control with different phases in Supplementary Section C.

\section*{Acknowledgements}

AH gratefully acknowledges the sponsorship from the City University of Hong Kong (Project No. 7005615); and the Hong Kong Institute for Advanced Study, City University of Hong Kong (Project No. 9360157). The work described in this paper was substantially supported by a grant from the Research Grants Council of the Hong Kong Special Administrative Region, China (Project No. CityU 11200120).

\section*{Author contributions statement}

Y.C. designed and implemented the quantum algorithms. All authors contributed to interpreting data and engaged in useful scientific discussions. All authors reviewed the manuscript.

\section*{Additional information}

To include, in this order: \textbf{Accession codes} (where applicable); \textbf{Competing interests} (mandatoery statement). 

The corresponding author is responsible for submitting a \href{http://www.nature.com/srep/policies/index.html#competing}{competing interests statement} on behalf of all authors of the paper. This statement must be included in the submitted article file.
\end{document}


\title{The quantum imaginary-time control for accelerating the ground-state preparation}

\author{Yu-Cheng Chen}
\affiliation{Tencent Quantum Laboratory, Tencent, Shenzhen, Guangdong, China, 518057}
\affiliation{Department of Mechanical Engineering, City University of Hong Kong, Hong Kong SAR, 999077, China}
\author{Yu-Qin Chen}
\email{yuqinchen@tencent.com}
\affiliation{Tencent Quantum Laboratory, Tencent, Shenzhen, Guangdong, China, 518057}
\author{Alice Hu}
\email{alicehu@cityu.edu.hk}
\affiliation{Department of Mechanical Engineering, City University of Hong Kong, Hong Kong SAR, 999077, China}
\affiliation{Department of Materials Science and Engineering, City University of Hong Kong, Kowloon, Hong Kong SAR, 999077, China}
\author{Chang-Yu Hsieh}
\email{kimhsieh2@gmail.com}
\affiliation{Tencent Quantum Laboratory, Tencent, Shenzhen, Guangdong, China, 518057}
\affiliation{{Innovation Institute for Artificial Intelligence in Medicine of Zhejiang University, College of Pharmaceutical Sciences, Zhejiang University, Hangzhou, 310058, China}}
\author{Shengyu Zhang}
\affiliation{Tencent Quantum Laboratory, Tencent, Shenzhen, Guangdong, China, 518057}
\maketitle
\onecolumngrid
\appendix

\tableofcontents

\section{Acceleration Induced by Lyapunov Control} \label{apptheory}
In this section, we explain the simulation acceleration driven by Lyapunov control in multiple stages. First, we discuss a practical strategy to construct a control Hamiltonian (without explicitly diagonalizing the system Hamiltonian) that ensures substantial speedup in section \ref{appgeneralcontrol}. Second, we attempt to provide an intuitive understanding of the accelerated dynamical process by inspecting time-dependent changes in the distribution of eigenstate populations and the time-evolved energy spectrum of $H(\tau)$ in section \ref{appdynamicpross}. In section \ref{appcontrol}, we look at a broad spectrum of realistic models (of interest in condensed matter physics, quantum chemistry, and combinatorial optimization) to demonstrate the broad applicability of the proposed method for simulating the ground state of a Hamiltonian.

\subsection{General Strategy For Control Hamiltonian Construction} \label{appgeneralcontrol}

As stated in the previous section, one scenario to drive the desired exponential speedup is to construct an $H_d$ that (1) commutes with $H_p$, (2) maintains the eigenstate ordering of $H_p$ and (3) holds the matrix norm for $H_p + H_d$ roughly the same as that of $H_p$. Among the three criteria, it is reasonable to partially relax the condition on the eigenstate ordering and make $H_d(t)$ a time-dependent control Hamiltonian instead.

One possibility is to choose $H_d(t)=\sum_{i=1}^n \beta_i(t) H_p^i$, which is an $n$-th order matrix polynomial made up of power of $H_p^i$ and $\beta_i \in \mathbf{R}$. Intuitively, the time-dependent spectrum of $H_d(t)$ may temporarily alter eigenstate ordering and enlarge or shrink energy gaps between states. Under this temporal modulation of the Hamiltonian spectrum, there would be accelerated and de-accelerated population transfer among eigenstates of $H_p$ during the imaginary time evolution. To ensure that we obtain as large an energy filtering towards the ground state of $H_p$ as possible is to simply minimize the expected energy value $\bra{\psi(t)}H_p\ket{\psi(t)}$ during the imaginary time evolution under $H_p+H_d(t)$. The gradient descent will give us the rule to update $\beta(t)$. At this point, this strategy of optimizing $H_d(t)$ essentially corresponds to the closed-loop control theory discussed thoroughly in the main text.

We now have a simple and clear picture for the selection of control Hamiltonian. However, it might not be experimentally feasible to implement such a complex form of control Hamiltonian, $H_d(t)=\sum_{i=1}^n \beta_i(t) H_p^i$. An approximate option is to retain only the major components of $H_p$ in this polynomial expansion of $H_d(t)$. Effectively, instead of having an $H_d(t)$ that always commutes with $H_p$, we end up considering an $\tilde H_d(t)$ with non-zero off-diagonal matrix elements in the eigenbasis of $H_p$. While this compromise (due to realistic experimental constraints) may offset the exponential speedup, we find that it is still possible to achieve an appreciable speedup within the closed-loop control setup. 
To make the discussion in this section self-contained, we present a numerical illustration.

\begin{figure*}
\begin{center}
\hspace*{-0cm}\includegraphics[width=0.8\textwidth]{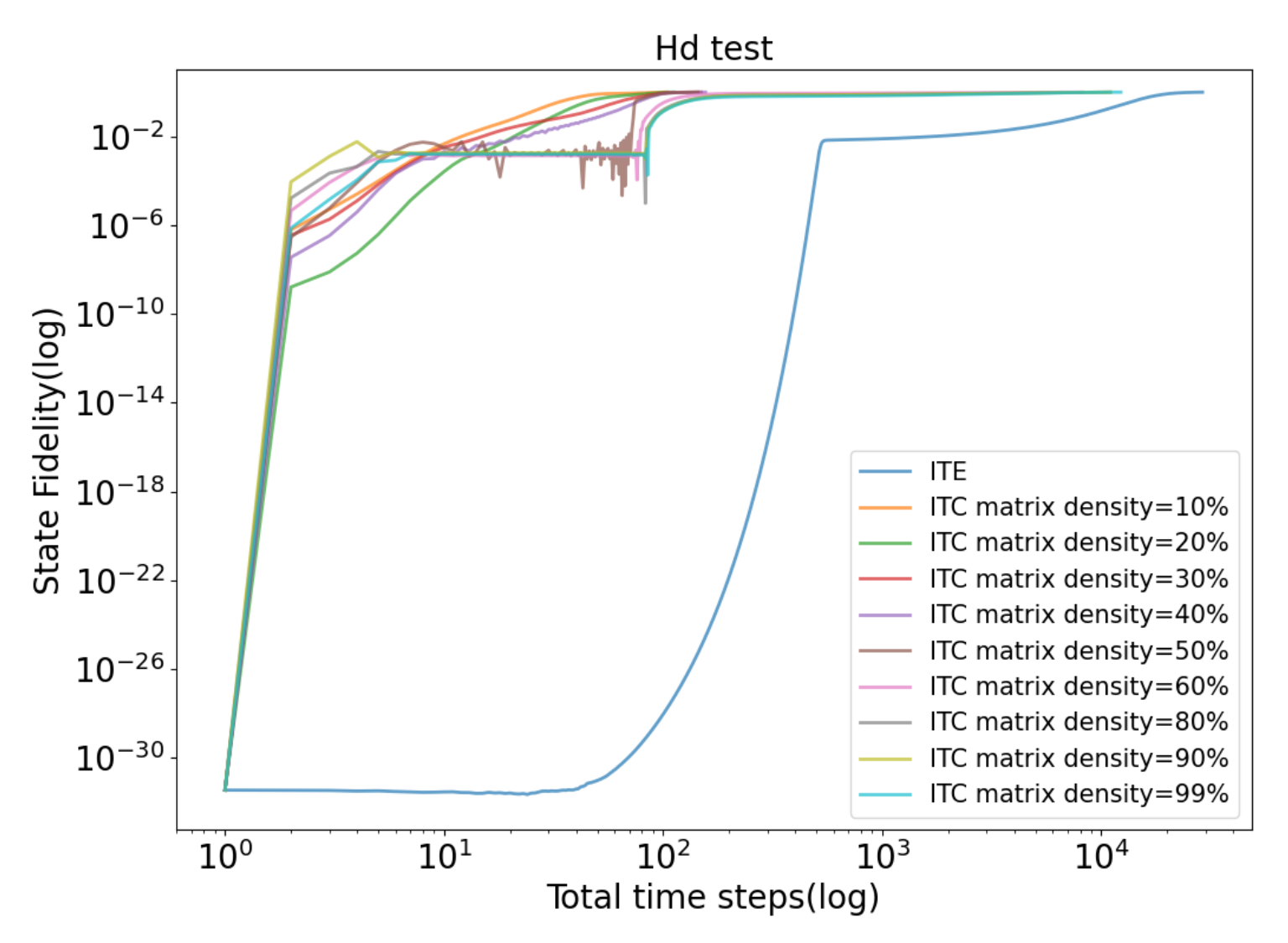}
\caption{The convergence results of $H_d$. The ITC matrix density is the percentage of the matrix $\tilde{D}$ filled with the nonzero number range from 0 to 1. The x-axis is the log of the total convergence time step. The y-axis is the fidelity between the convergence state and the ground state.
}\label{Dtest}
\end{center}
\end{figure*}

In Figure \ref{Dtest} we show the simulation result of how the imperfectly controlled dynamics, driven by $H_p + \tilde H_d(t) =  U \tilde{D}(t) U^\dag$, affect the convergence of imaginary-time evolution for a  9 qubits 2D XXX model with non-periodic boundary condition. The problem Hamiltonian can be written as $H_p=0.2\sum_i Z^i+0.1 \sum_{edge}X^iX^j+Y^iY^j+Z^iZ^j$, where edge implies the nearest-neighbor couplings and the initial state is $\ket{++\dots+}$. Assume that the originally intended control Hamiltonian $H_d$ is intended to enlarge the energy gaps between the ground state and excited states, but, due to the experimental constraints, we choose to adopt a more easily implementable $\tilde H_d$. In this case, we no longer have a fully diagonal matrix $D$ in the eigenbasis of $H_p$ rather we consider 
$$\tilde{D}=\begin{pmatrix}
-5& 0 & 0 & ...\\
0 & 5 & 0 & ...\\
... & ... & ... & ...\\
0 & 0 & 0 & 0
\end{pmatrix}+\mathcal{R}(p),$$
where $\mathcal{R}(p)$ is a random sparse matrix with $p$ percentage of non-zero off-diagonal matrix elements. The result shows that the $H_d$ with non-trivial off-diagonal disorder $\tilde{D}(p=10\%)$ can still maintain the exponential speed-up for this problem, although the further increase of the number of off-diagonal terms will eventually suppress the speedup. Interestingly, when the number of non-zero off-diagonal disorders rises up to $p=60\%$ the inhibition effects on the speedup of this 9-qubit system seem to plateau. 

In a practical scenario, the control Hamiltonian for the models in Appendix \ref{appcontrol} follows the rules we mentioned above. For molecular systems studied, Pauli Z terms are the dominant terms in $H_p$ with large coefficients. The selection of  those dominant Pauli $Z$ terms as control Hamiltonian follows the rule (1) since it commutes with the highest contribute terms in $H_p$ therefore approximately commutes with $H_p$. The details of $H_d$ selection are listed in Appendix \ref{appmoleculecontrol}. For complex systems like the XXX model and SK model, Pauli $Z$ terms from $H_p$ may not be dominant parts. We select the $k$-local Pauli $Z$ as $H_d$, which not only commutes with all Pauli $Z$ terms but also commutes with many Pauli XX(YY) terms making it approximately follow the rule (1). Finally, proper design of the control strategy and maximum admissible control impulses will allow rules (2) and (3) to be successfully followed, see details in \ref{appstrategy}.

\subsection{Dynamic process visualization} \label{appdynamicpross}

\begin{figure}[ht]
\begin{center}
\includegraphics[width=0.7\textwidth,height=0.4\textwidth]{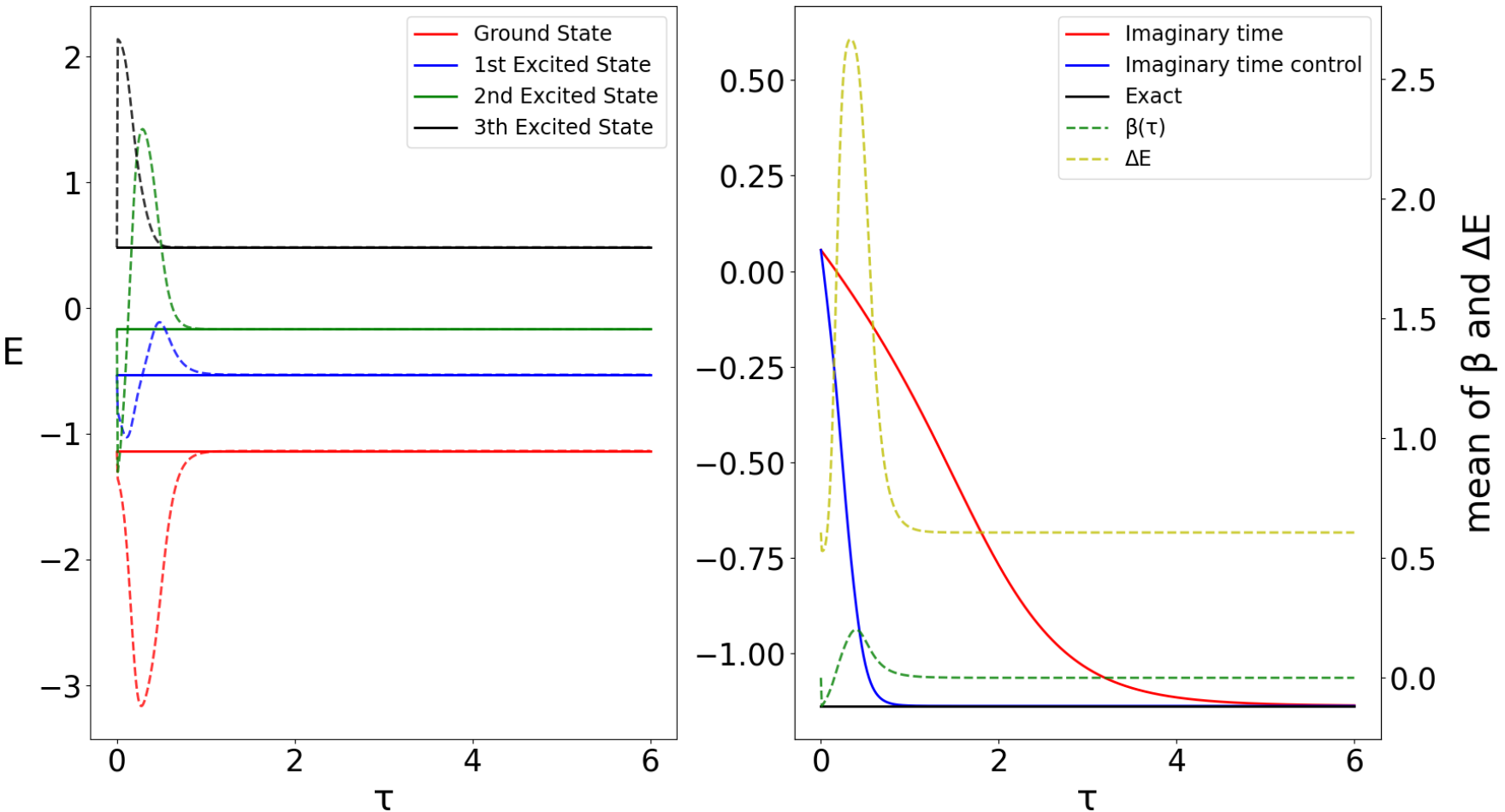}
\caption{The result of 2-qubit hydrogen with complete control. The left figure shows how the energy levels change with time, as the solid line is the original energy level and the dashed line is the controlled energy level. The right figure is the convergence of imaginary time models. The left y-axis corresponds to the energy for the red solid line and blue solid line. The right y-axis corresponds to the energy gap between the ground state and the first excited state, and the $\beta(\tau)=\frac{1}{n}\sum_{i=1}^n\beta_i(\tau)$. The result shows that all energy levels will change with time.}\label{FULLH2}
\end{center}
\end{figure}

\begin{figure}[ht]
\begin{center}
\includegraphics[width=0.7\textwidth,height=0.4\textwidth]{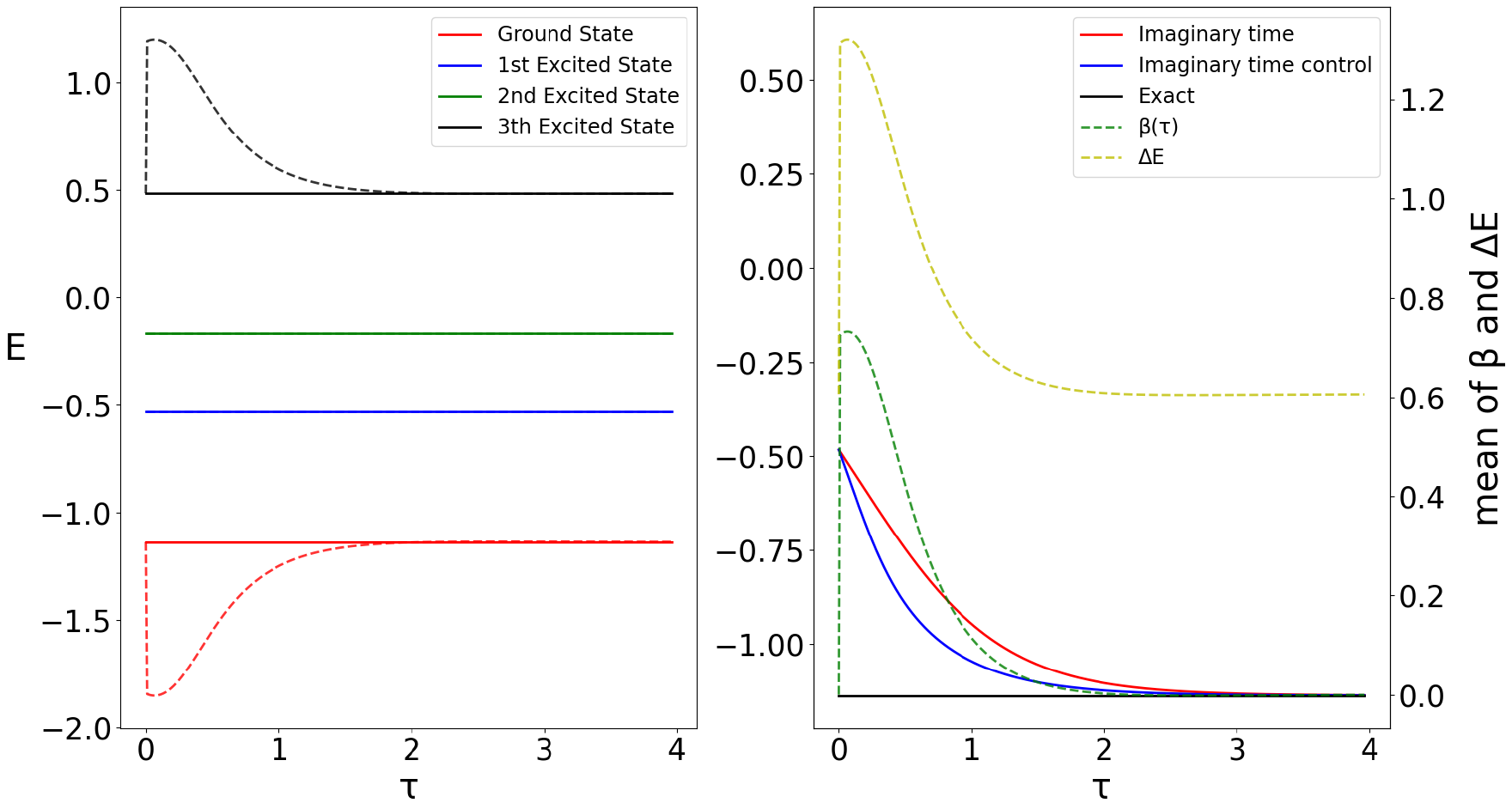}
\caption{The result of 2-qubit hydrogen with non-complete control shows that the highest excited state and the ground state will both change with time.}\label{PARTH2}
\end{center}
\end{figure}

We find the ITC temporally modulates the entire spectrum of $H(\tau)$ during the evolution and return back to the original spectrum of $H_p$ when time $t>>1$ as $\beta(t>>1)\rightarrow 0$. As can see in the following figures, for the case of the imaginary time evolution (ITE), since the $H_p$ is a time-independent Hamiltonian, the energy gap between the ground state and the excited states remains fixed in time. For the imaginary time control (ITC), on the other hand, the energy gap between the ground state and the excited states of $H_p$ will change with time. However, unlike the polynomial basis that can guarantee to order of the whole eigenspectrum as in Figure \ref{FULLH2}, the small set of limited Pauli basis might only modify some of the eigenstates as in Figure \ref{PARTH2}. We comment on how the energy gap changes with time under control using different control Hamiltonians from the point of view of controllability, which provides another realization of how the control Hamiltonian orders the energy spectrum during the convergence. The idea is borrowed from the idea of controllability discussed in real-time control \cite{Zhang2005}, in our definition, given a set of reachable states denoted by $\mathcal{R}(\ket{\psi_0})$ start from initial state $\ket{\psi_0}$, the control system $H_p+\beta(\tau) H_d$ is said to be completely (eigenstates) controllable if $\mathcal{R}(\ket{\psi_0})=\mathcal{E}$, where $\mathcal{E}$ is the collection of eigenstates, and said to be non-completely (eigenstates) controllable if $\mathcal{R}(\ket{\psi_0})\neq\mathcal{E}$. From the point of view of the energy spectrum, since the final convergence state of imaginary time evolution should be the ground state of the new system, the control system should reorder the whole energy spectrum during the convergence when the control system is completely controllable. For the control system that is not completely controllable, it can still accelerate the ground state convergence if the ground state in the reachable state $\mathcal{R}(\ket{\psi_0})$ which should always be true since ITC can switch back to the original ITE by the turn of the $\beta$. By definition, it is clear that polynomial basis is in the class of completely controllable, and limited Pauli basis might be in one of the classes according to the elements in the basis. In the following numerical experiment, we will visualize the concept of controllability using the energy spectrum.
The energy gap between the ground state and the excited states of $H_p$ will change with time as,
$$
\Delta E(\tau) = \expval{H(\tau)}{\psi_0}-\expval{H(\tau)}{\psi_i}.
$$
where $H(\tau)=H_p+\beta(\tau)H_d$, $\ket{\psi_0}$ and $\ket{\psi_i}$ correspond to the ground state and i-th excited state of drift Hamiltonian $H_p$, respectively. In Fig.~\ref{FULLH2} we show the simulated result of using an ITC to prepare the ground state for a 2-qubit hydrogen model. In this case, we adopt a set of control Hamiltonians that yield complete controllability (polynomial basis), on the 2-qubit dynamics. As clearly illustrated in the figure, the ITC temporally modulates the entire spectrum of $H(\tau)$ during the evolution and returns back to the original spectrum of $H_p$ when time $t>>1$ as $\beta(t>>1)\rightarrow 0$.  Thus, we conclude that the controlled system can evolve towards the ground state of $H_p$ in fewer time steps with enlarged energy gaps.

Next, we attempt a different experiment. If the control Hamiltonian does not guarantee complete controllability(limited Pauli basis), the imaginary-time control will not be able to enlarge all energy gaps but just a few of them. In this case, it can still speed up the convergence of the ITC but may not be as fast as the previous case with complete controllability In Fig.~\ref{PARTH2}, the 2-qubit Hydrogen is driven by a non-complete controllable Hamiltonian $H_d$, the instantaneous ground state energy drop increases the energy gap between the ground state and the excited states, therefore, provide the speed-up for ITC, and this control will benefit more when the initial state has an appreciable overlap with the highest excited state since the highest excited state is also controllable. 
\par

\par
\par
\begin{figure}[ht]
\begin{center}
\includegraphics[width=0.7\textwidth,height=0.4\textwidth]{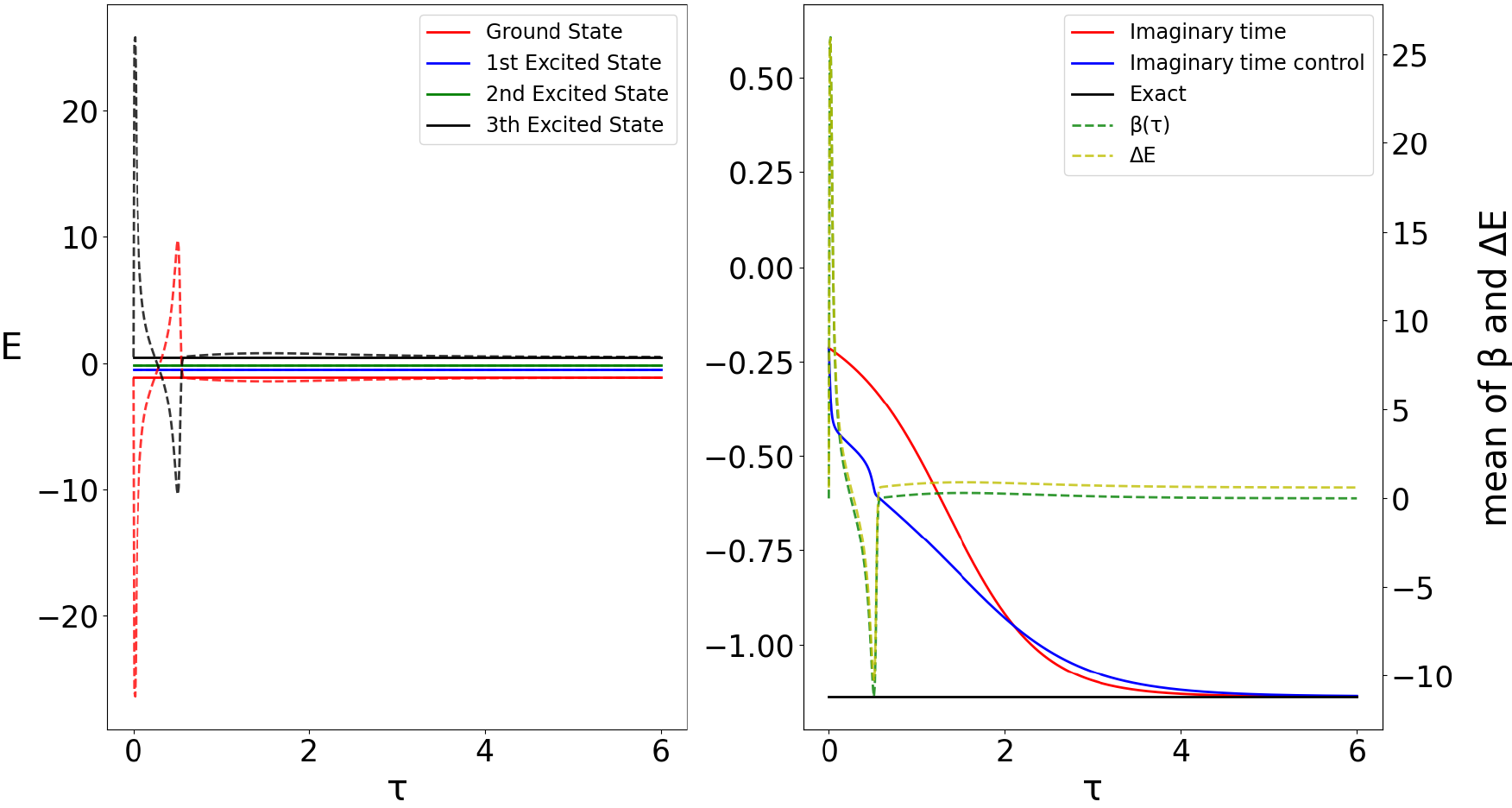}
\caption{The result of 2-qubit hydrogen with aggressive control strategy shows that the control pulses are able to rearrange energy level.}\label{badH2}
\end{center}
\end{figure}

However, when the control Hamiltonian is not completely controllable and the control strategy $\beta(\tau)$ is too aggressive, then the driven ITE may actually compromise the rate of convergence. This is because the ground state has been inverted to the highest excited state in some time steps, and the system will be temporarily driven away from the target state (i.e. the ground state of $H_p$) when its energy has been shifted upwards. In Fig.~\ref{badH2}, during the
period $\tau \in (0.5,1.0)$, the order of the ground state and the highest excited state are swapped. Clearly, as shown in the right panel of Fig.~\ref{badH2}, the ITE converges to the desired ground state faster in this case.

\begin{figure}[ht]
\begin{center}
\includegraphics[width=0.7\textwidth,height=0.4\textwidth]{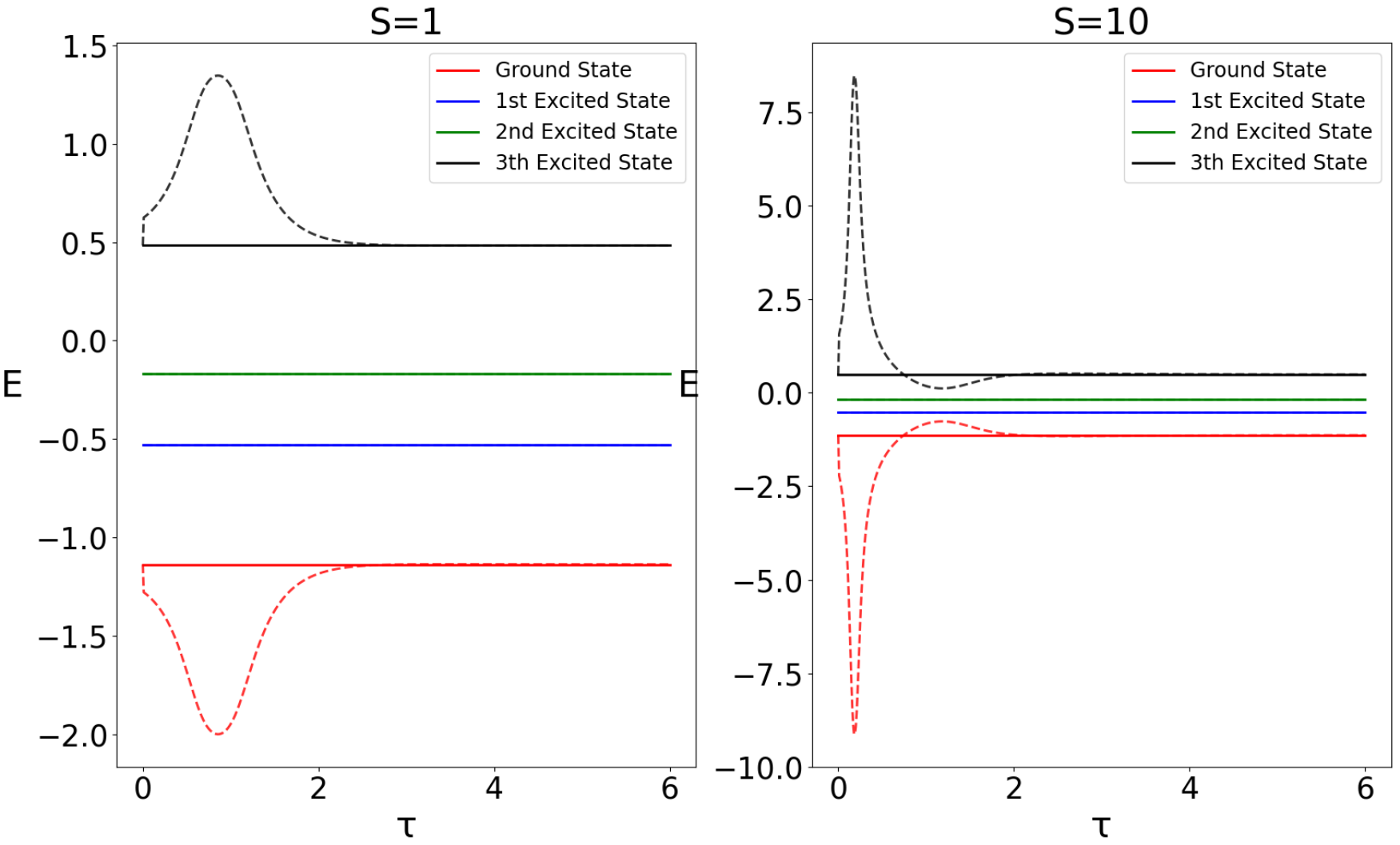}
\caption{The result of 2-qubit hydrogen with different admissible maximum strength S. It shows that the large admissible maximum strength might decrease the energy gap.}\label{S1VS10}
\end{center}
\end{figure}

To avoid the unintended eigenstate re-ordering by the control field, it is crucial to confine the magnitude of $\beta$ to some proper range. In our test, it is better to set $\vert \beta \vert$ to be equal to or less than the magnitude of the energy $\bra{\psi(\tau)}H_p \ket{\psi(\tau)}$ which can be obtained from measuring the norm of the matrix problem Hamiltonian. In Fig.~\ref{S1VS10} we illustrate the effects of choosing different $\beta$ on the time-evolved energy levels of $H_p$, which is the Hamiltonian for a 2-qubit Hydrogen. In this test, we adopt two different approximate bang-bang control strategies $\beta(\tau)=\frac{2S}{1+e^{-\gamma T_k}}-S$, introduced in appendix \ref{appstrategy}. These two control strategies share the same $\gamma$ but $S=1$ and $S=10$, respectively. From Fig.~\ref{S1VS10}, it is clear that bigger $\vert \beta\vert$ may enlarge energy gaps between eigenstates of $H_P$ but also increases the likelihood of state re-ordering.

\section{Empirical Control Hamiltonian Selection} \label{appcontrol}
As mentioned in Appendix \ref{apptheory}, the success of the imaginary time control (ITC) is to modify this typical evolution path of ITE. The control may drive the time-evolved wavefunction to temporarily enhance contributions from high-energy eigenstates before converging to the ground state. In other words, the advantage of having controlled dynamics in the imaginary-time domain is to avoid densely spaced energy regions in the Hamiltonian spectrum, which cannot be escaped in the standard ITE. If one can find a $H_d$ that could drive a system towards high-energy states during the ITE, it has the potential to improve the convergence efficiency as prescribed by our proposed method to get ITC. 

In this section,  we present the strategies of control Hamiltonian selection and discuss how different control Hamiltonian impact the 
the execution of quantum computing experiments, including empirically control Hamiltonian and simulation results for molecule in Appendix \ref{appmoleculecontrol}, control Hamiltonian, and control Hamiltonian and simulation results for Sherrington–Kirkpatrick in Appendix \ref{appskcontrol}, control Hamiltonian and simulation results for a spin model constructed from 3-SAT in Appendix \ref{appsatcontrol}.
In summary, with a proper design of control Hamiltonian, we will obtain obvious acceleration for  ground state simulation. Thus our proposed imaginary-time control can potentially make a digital quantum simulation for the ground-state preparation much more accessible in the NISQ era.

\subsection{Control Hamiltonian and simulation results for molecule system} \label{appmoleculecontrol}
Based on the above general strategy for control Hamiltonian selection, here we provide a further empirical method for control Hamiltonian selection, especially for molecule systems. We find that the empirical $H_d$ and its slightly modified versions can exponentially speed up the ITC convergence for studied systems with both large and small energy gaps.

For the molecule system, we try the candidate control Hamiltonian $H_i$ containing only Pauli $Z$ (single $Z$ and double $Z$, and the total number of choices is $C_1^n+C_2^n$ where $n$ is a number of qubits. In studied molecular cases, we find that limiting the Pauli $Z$ to quadratic is enough to obtain exponential speed-up. 

Intra-molecular bond distance for the typical dissociation curves is considered as an adjustable parameter to generate variant energy gaps for the molecular systems. We choose a problem Hamiltonian $H_p$ with a large energy gap as a fast test to help with control Hamiltonian selection. We first run standard ITE on this $H_p$ (which should converge to the ground state easily owing to the large energy gap) and keep a record of the $\beta_i(\tau)$ from the ITE calculation with $\ket{+++ \dots ++}$ as the initial state. We sum up $\beta_i$ over time $B_i=\sum_{\tau} \beta_i(\tau)$ for further analysis. From our numerical investigations on molecular systems: H chain, LiH, and HF, we find that those $H_i$ having negative $B_i$ all share the same structure. The structures of Hamiltonian means that, the Pauli strings are divided into two groups that occupy $[0:(N-1)]$ and $[N:(M-1)]$ orbitals respectively, where $N$ is the number of electrons and $M$ is the number of total orbitals. We use these $H_i$ as the control Hamiltonian, which we called \emph{Empirical $H_d$}. Empirical $H_d$ and its slightly modified versions can exponentially speed up the ITC convergence for studied systems with both large and small energy gaps.

 \begin{table}[!htb]
\begin{tabular}{|l|c|c|c|c|c|c|}
\hline
 &
  \multicolumn{1}{l|}{$H_2$} &
  \multicolumn{1}{l|}{$H_4$} &
  \multicolumn{1}{l|}{$H_6$} &
  \multicolumn{1}{l|}{$H_8$} &
  \multicolumn{1}{l|}{LiH} &
  \multicolumn{1}{l|}{HF} \\ \hline
\multirow{4}{*}{Full $H_d$} &
  II+P(IZ) &
  IIII+P(IIIZ) &
  IIIIII+P(IIIIIZ) &
  IIIIIIII+P(IIIIIIIZ) &
  IIII+P(IIIIIIIZ) &
  IIIIIIIIII+P(IZ) \\ \cline{2-7} 
 & P(IZ)+II & P(IIIZ)+IIII & P(IIIIIZ)+IIIIII & P(IIIIIIIZ)+IIIIIIII & P(IIIZ)+IIIIIIII & P(IIIIIIIIIZ)+II \\ \cline{2-7} 
 & II+P(ZZ) & IIII+P(IIZZ) & IIIIII+P(IIIIZZ) & IIIIIIII+P(IIIIIIZZ) & IIII+P(IIIIIIZZ) & IIIIIIIIII+P(ZZ) \\ \cline{2-7} 
 & P(ZZ)+II & P(IIZZ)+IIII & P(IIIIZZ)+IIIIII & P(IIIIIIZZ)+IIIIIIII & P(IIZZ)+IIIIIIII & P(IIIIIIIIZZ)+II \\ \hline
\multirow{2}{*}{Half $H_d$} &
  II+P(IZ) &
  IIII+P(IIIZ) &
  IIIIII+P(IIIIIZ) &
  IIIIIIII+P(IIIIIIIZ) &
  IIII+P(IIIIIIIZ) &
  IIIIIIIIII+P(IZ) \\ \cline{2-7} 
 & II+P(ZZ) & IIII+P(IIZZ) & IIIIII+P(IIIIZZ) & IIIIIIII+P(IIIIIIZZ) & IIII+P(IIIIIIZZ) & IIIIIIIIII+P(ZZ) \\ \hline
 \multirow{2}{*}{All $H_d$} &
  P(IIIZ) &
  P(IIIIIIIZ) &
  P(IIIIIIIIIIIZ) &
  P(IIIIIIIIIIIIIIIZ) &
  P(IIIIIIIIIIIZ) &
  P(IIIIIIIIIIIZ) \\ \cline{2-7} 
 & P(IIZZ) & P(IIIIIIZZ) & P(IIIIIIIIIIZZ) & P(IIIIIIIIIIIIIIZZ) & P(IIIIIIIIIIZZ) & P(IIIIIIIIIIZZ) \\ \hline
\multirow{3}{*}{Empirical $H_d$} &
  II+P(IZ) &
  IIII+P(IIIZ) &
  IIIIII+P(IIIIIZ) &
  IIIIIIII+P(IIIIIIIZ) &
  IIII+P(IIIIIIIZ) &
  IIIIIIIIII+P(IZ) \\ \cline{2-7} 
 & II+P(ZZ) & IIII+P(IIZZ) & IIIIII+P(IIIIZZ) & IIIIIIII+P(IIIIIIZZ) & IIII+P(IIIIIIZZ) & IIIIIIIIII+P(ZZ) \\ \cline{2-7} 
 & P(ZZ)+II & P(IIZZ)+IIII & P(IIIIZZ)+IIIIII & P(IIIIIIZZ)+IIIIIIII & P(IIZZ)+IIIIIIII & P(IIIIIIIIZZ)+II \\ \hline
\end{tabular}
\caption{The $H_d$ selection of figures. The function P(S) is the collection of all permutations without repetition of Pauli string S. The I+P(S) means add I before all the Pauli string P(S), for example, II+P(IZ) is equal to \{IIIZ, IIZI\}. The Full $H_d$ is slightly different from empirical $H_d$ since it has the P(IZ)+I term and the Half $H_d$ is slightly different from the empirical $H_d$  since it lacks the P(ZZ)+I term.
}\label{T1}
\end{table}
Take it for an example, there are 4 electrons ($N=4$) and 12 total orbitals ($M=12$), the Pauli strings $P_0P_1P_2P_3P_4P_5P_6P_7P_8P_9P_{10}P_{11}$ will be divided into two groups: $P_0 P_1 P_2 P_3 IIIIIIII$ and $IIIIP_4 P_5 P_6 P_7 P_8 P_9 P_{10} P_{11}$ (for Openfermion notation, the energy of the orbitals from low to high are ordered from left to right). We find that the single $Z$ Pauli strings that $Z$ only exist in $[N:(M-1)]$ orbitals and the double $Z$ Pauli strings that $Z$s exist only in $[0:(N-1)]$ orbitals or $[N:(M-1)]$ orbitals have negative $B_i$ (as Table \ref{T1} ``Empirical $H_d$"), and all other Pauli strings have positive $B_i$. 

\begin{figure*}
\begin{center}
\hspace*{-0cm}\includegraphics[width=1\textwidth]{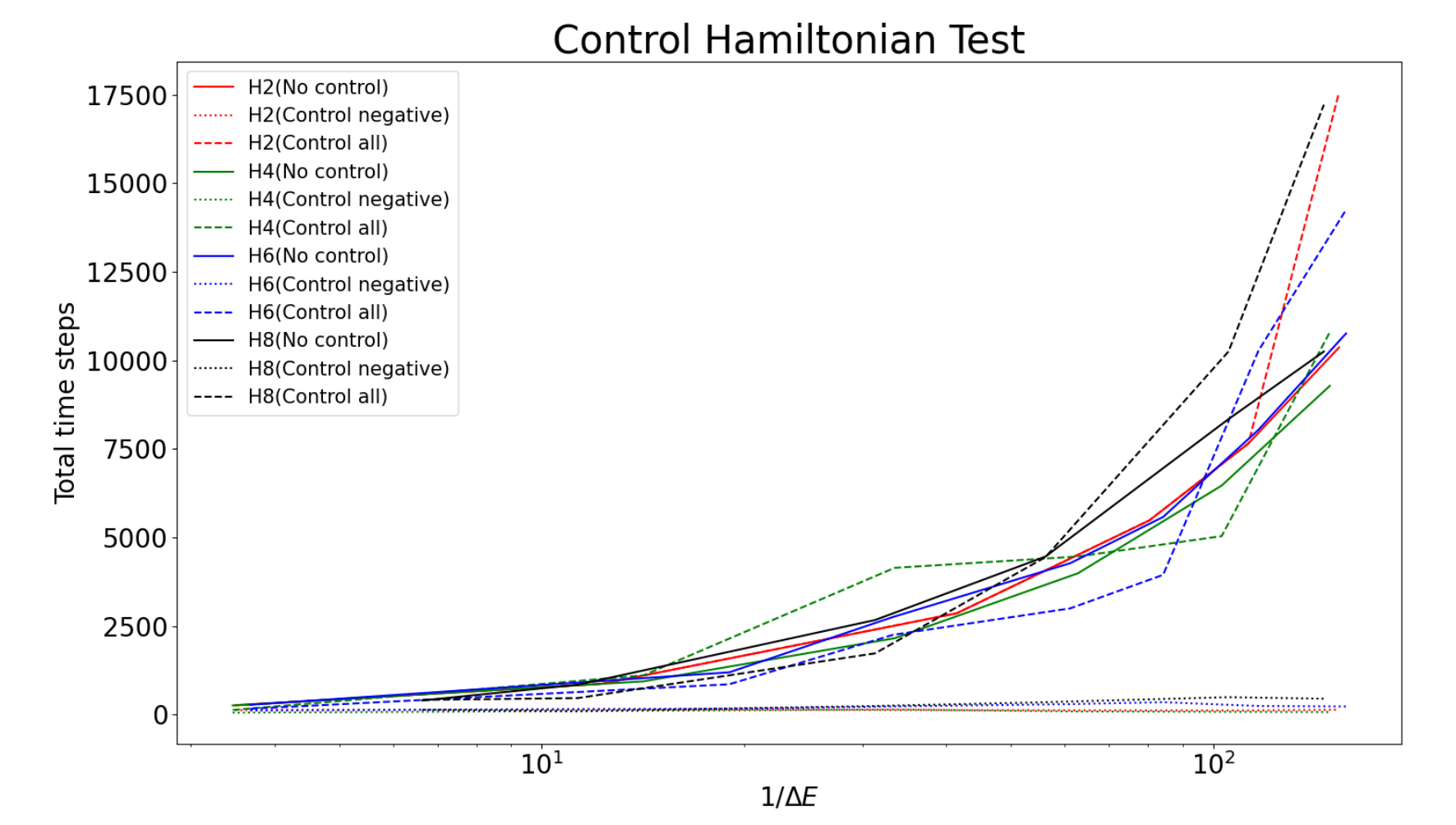}
\caption{The result of the control Hamiltonian test of the H chain system, the x-axis is a log of 1/$\Delta E$, and the y-axis is the total time steps for convergence. The result shows that the empirical control Hamiltonian can provide an ITC that gives exponential speed-up with respect to the standard ITE for all cases while $H_d$ that contain all $H_i$ require extra time or no speed-up.
}\label{HD1}
\end{center}
\end{figure*}

\begin{figure*}
\begin{center}
\hspace*{-0cm}\includegraphics[width=0.95\textwidth]{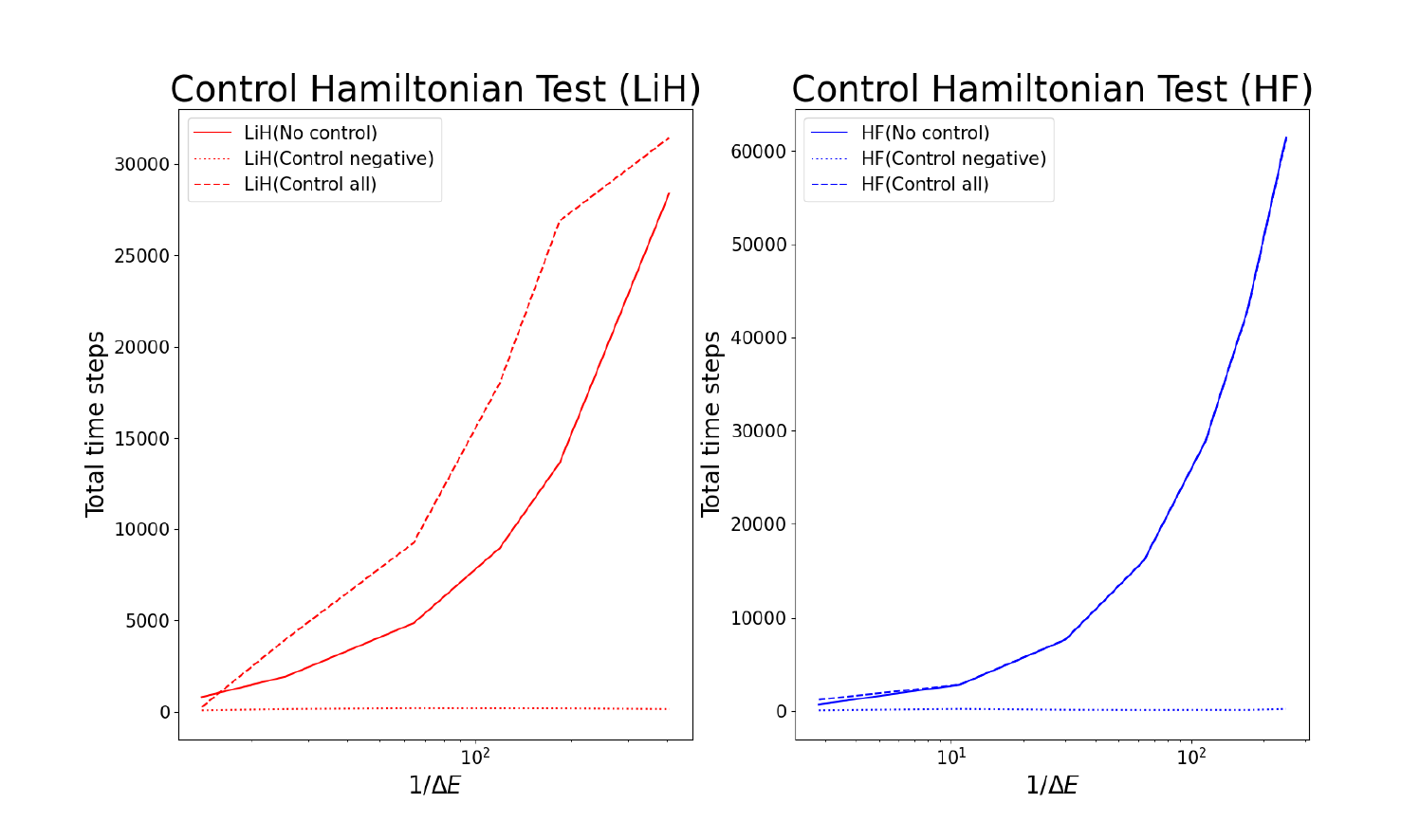}
\caption{The result of the control Hamiltonian test of LiH and HF, the x-axis is a log of 1/$\Delta E$, and the y-axis is the total time steps for convergence. The result shows that the empirical control Hamiltonian can provide good control results that have exponentially speed-up from ITE for all cases while $H_d$ that contain all $H_i$ require extra time or no speed-up.
}\label{HD2}
\end{center}
\end{figure*}

Finally, we use the empirical $H_d$ described above for various molecules (H chain, LiH, and HF, see Appendix \ref{appmolecularham}.) with different energy gaps. In Figure \ref{HD1} and Figure \ref{HD2}, we show the ITC results with the control Hamiltonian that has all single $Z$ and double $Z$s (``control all" in Figure \ref{HD1} and Figure \ref{HD2}, ``All Hd" in Table \ref{T1}), the empirical control Hamiltonian mentioned above (``control negative" in Figure \ref{HD1} and Figure \ref{HD2}, ``Empirical Hd" in Table \ref{T1}) and the result of the standard ITE ("no control" in Figure \ref{HD1} and Figure \ref{HD2}). The results show that the empirical $H_d$ is far better than using all $H_i$ as control Hamiltonian. And empirical $H_d$ can exponentially speed up the ITC convergence for studied systems with both large and small energy gaps.

\begin{figure*}
\begin{center}
\hspace*{-0cm}\includegraphics[width=0.9\textwidth]{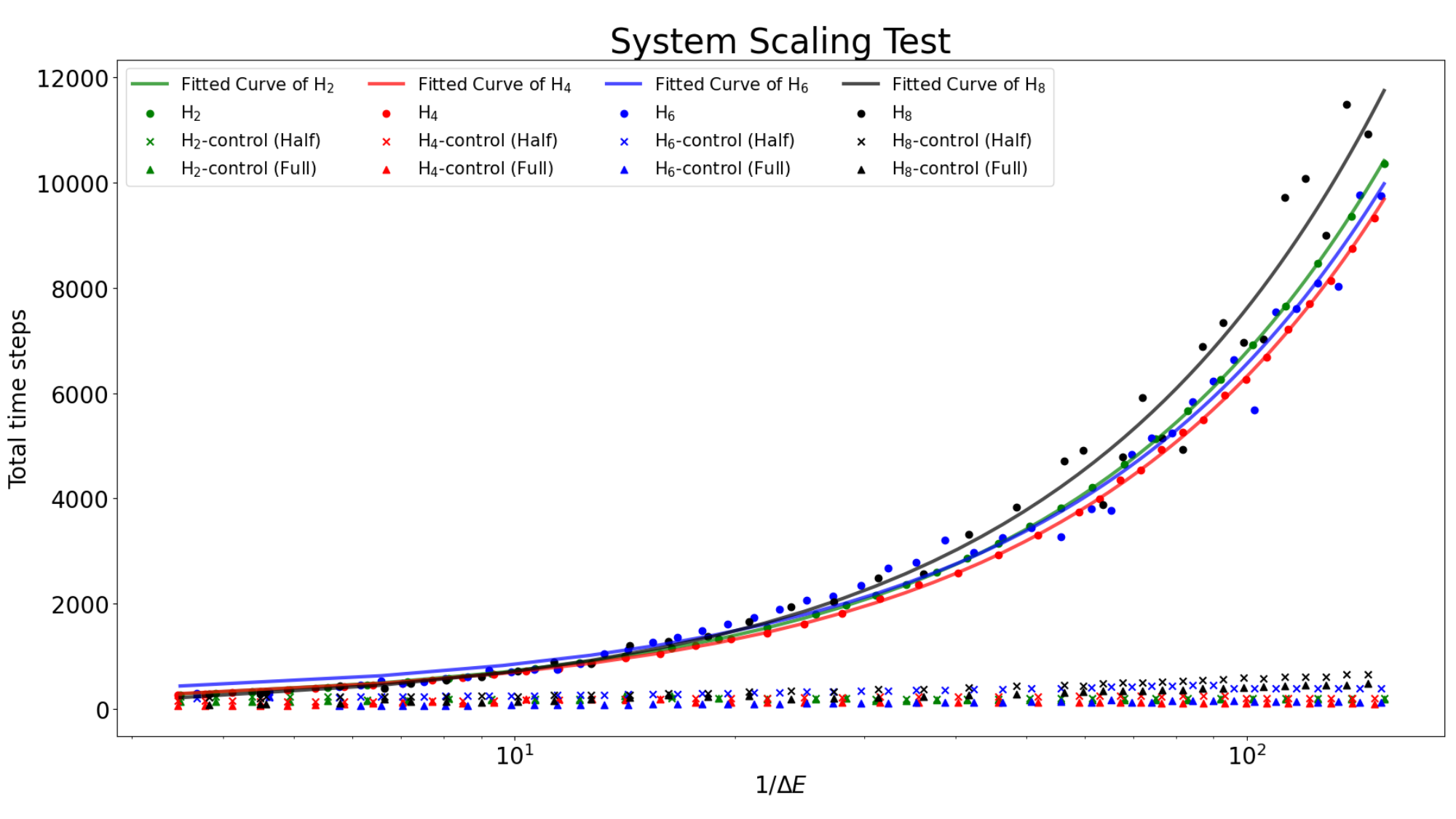}
\caption{The result of the H chain size scaling test with different types of $H_d$. The different dots represent different types of $H_d$ and different colors of lines represent the total time steps.  
}\label{Hlog}
\end{center}
\end{figure*}

We also test two types of $H_d$ that are slightly different from the empirical $H_d$ mentioned above. The details of those $H_d$ are also listed in Table \ref{T1}. Figure \ref{Hlog} shows the convergence of different types of $H_d$ selections and the speed-up for different system sizes, where the ``Full $H_d$" have extra single $Z$ terms compared to ``Empirical $H_d$" and the ``Half $H_d$" have less double Z terms compared to ``Empirical $H_d$". We can see that for ``Full $H_d$" and ``Half $H_d$" the acceleration scale exponentially as a function of $1/\Delta E$ for different choices of $H_p$. 

\subsection{Control Hamiltonian and simulation results for Sherrington–Kirkpatrick model} \label{appskcontrol}
Besides the commute basis $\{XX\dots XX, YY\dots YY, ZZ\dots ZZ\}$ we use in the 4 qubit cases in the main text. We also construct and compare two types of $H_d$ according to Appendix \ref{appgeneralcontrol}. One is constructed polynomial $H_d$ from the approximate polynomial basis $p(\tilde{H}_p)=\{\tilde{H}_p^2,\tilde{H}_p^3,\tilde{H}_p^4,\tilde{H}_p^5\})$, where approximate polynomial matrix $\tilde{H}_p^k$ is constructed from polynomial matrix $H_p^k$ by removing matrix elements that have values less than a fixed constant number. The other Pauli $H_d$ is selected from the 1-local Pauli Z and 4-local Pauli Z $cyclic(ZIIIIII)$ and $cyclic(ZZZZIIII)$, where cyclic means the set of circular shifts of Pauli strings, for example, $cyclic(XYZ)=\{XYZ, YZX, ZXY\}$.
The control strategy we use here is approximate bang-bang control with the maximum strength of the control field S slowly turning to zero within dozens of time steps to have lower resource requirement, see Appendix \ref{appstrategy}. 

\begin{figure}[ht]
\begin{center}
\includegraphics[width=0.7\textwidth]{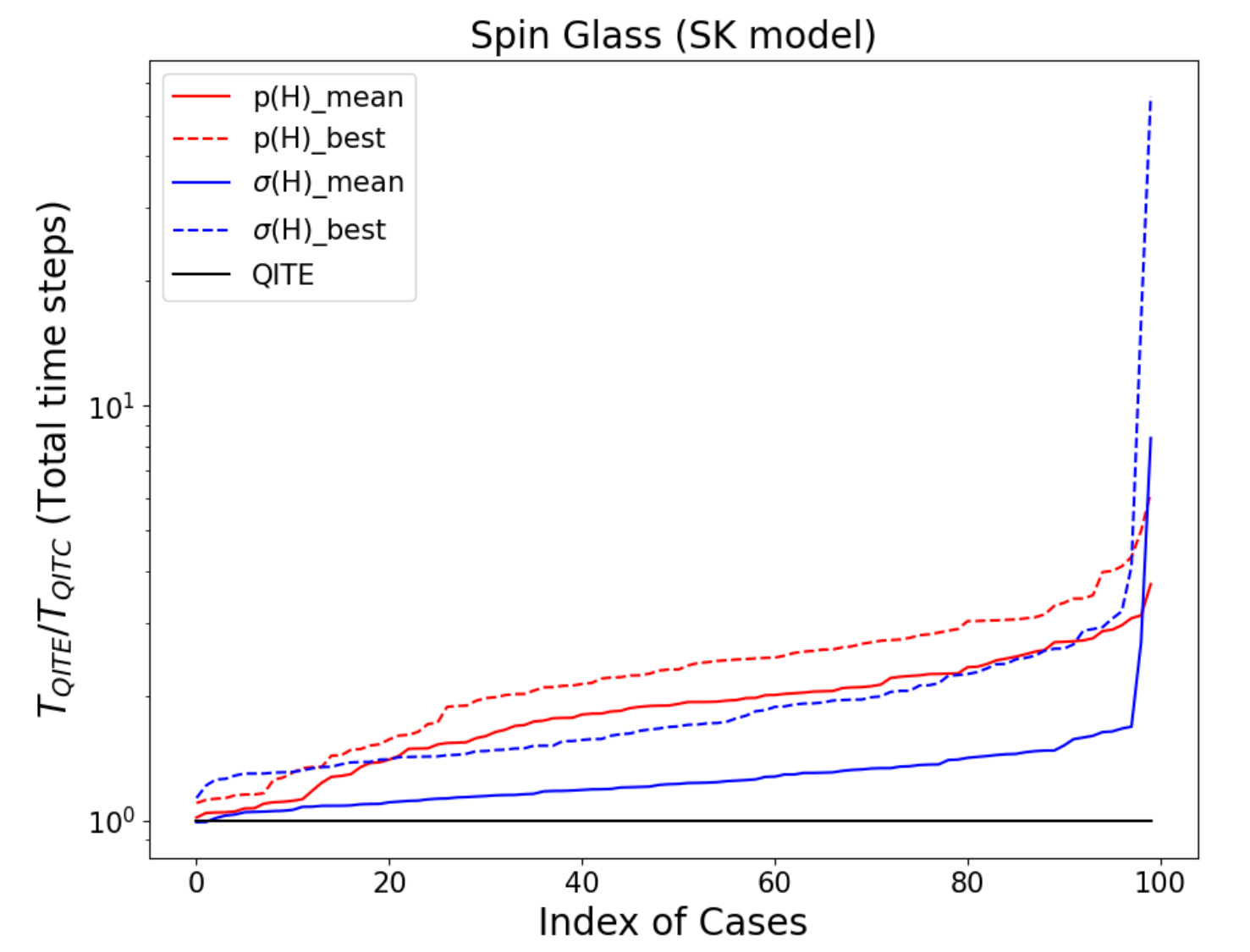}
\caption{The x-axis is the case indices sorted by the ratio of total QITE time steps divided by the total QITE time steps, note that sorting does not depend on energy differences. The y-axis is the ratio of the total time step. The mean of $p(H)$ and $\sigma(H)$ is the mean of the approximate polynomial $H_d$ and Pauli $H_d$ of ten randomly selected initial states. The best of $p(H)$ and $\sigma(H)$ is the largest speed-up among ten randomly selected initial states using approximate polynomial $H_d$ and Pauli $H_d$. The single time step $\Delta\tau=0.03$.}\label{SK}
\end{center}
\end{figure}
In Figure \ref{SK}, we compared the convergence of ITE and ITC of 100 randomly sampling cases and each case has 10 randomly selected initial states. The result shows that the control can provide a general speed-up and can achieve a hundred times speed-up in some cases. The approximate polynomial $H_d$ can provide a higher average speed-up but weaker corner case speed-up, while Pauli $H_d$ has a lower average speed-up but has a better corner case speed-up.

\FloatBarrier
\subsection{Control Hamiltonian and simulation results for a spin model constructed from 3-SAT} \label{appsatcontrol}
In the main text, we show that ITC can provide obvious speed-up over the ITE for molecular systems. It is also desirable to verify whether such a superior advantage can hold for other scenarios. In this subsection, we compare ITC and ITE for solving a spin model that is closely related to the 3-SAT problems. This is  another ground-state preparation task that is sufficiently distinct from the molecular systems considered in the main text. 

\begin{figure}[ht]
\begin{center}
\includegraphics[width=0.6\textwidth]{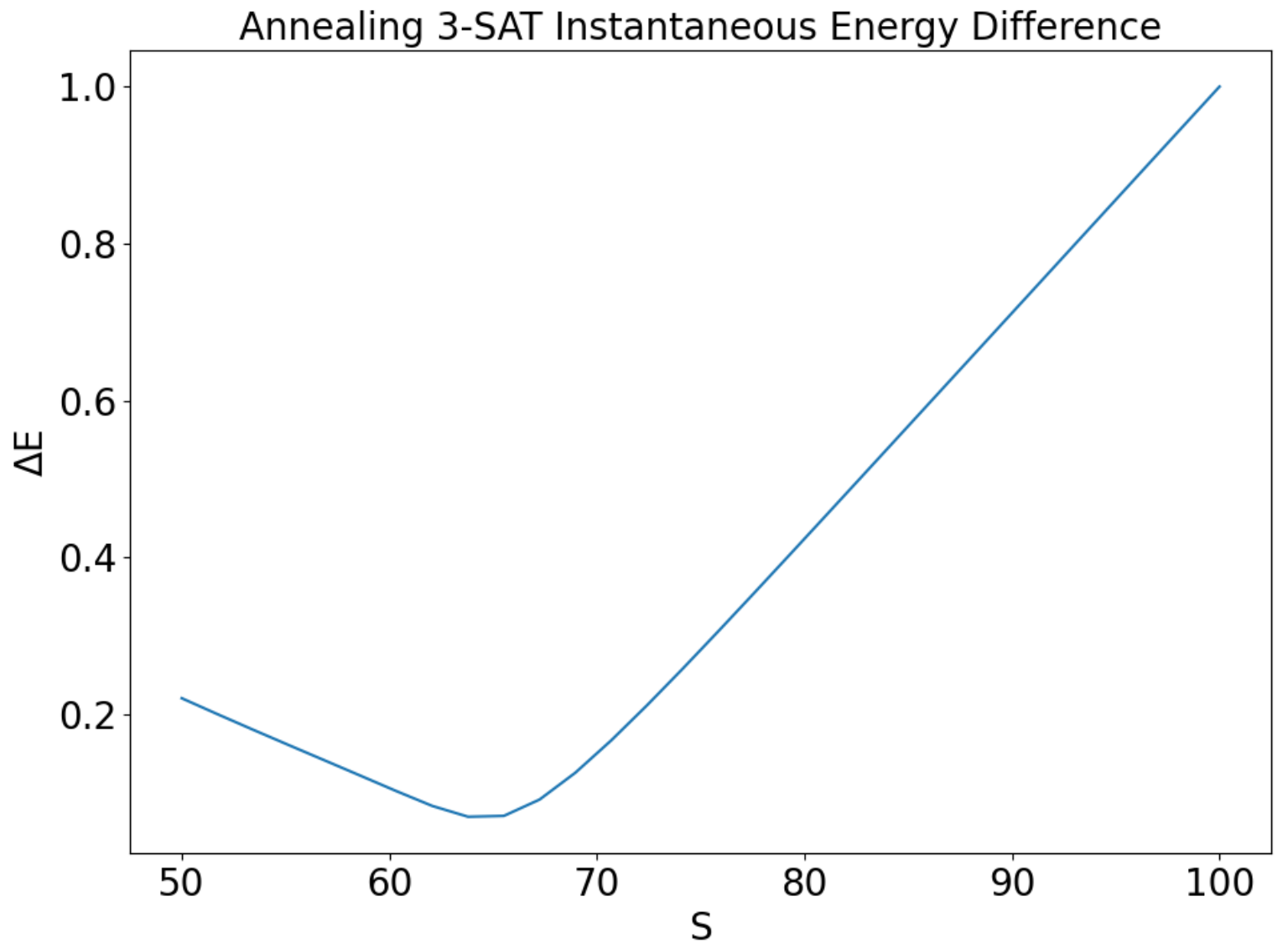}
\caption{The energy gap between the ground state and first excited state of the time-dependent Hamiltonian $H(s).$}\label{dEdS}
\end{center}
\end{figure}

3-SAT problem is defined by a logical statement involving $n$ boolean variables $b_i$. The logical statement consists of m clauses $C_i$ in conjunction:
$C_1 \wedge C_2 \wedge ... \wedge C_m$. Each clause is a disjunction of 3 literals, where a literal is a boolean variable $b_i$ or its negation $\lnot b_i$. For instance, a clause may read $(b_j \lor b_k \lor b_l).$ The task is to first decide whether a given 3-SAT problem is satisfiable; if so, then assign appropriate binary values to satisfy the logical statement. We can map a 3-SAT problem to a Hamiltonian for a set of qubits. Under this mapping, each binary variable $b_i$ is represented as a qubit state. Thus, an n-variable 3-SAT problem is mapped into a Hilbert space of dimension $N = 2^n$. Furthermore, each clause of the logical statement is translated to a projector acting on this n-qubit system. Hence, a logical statement with m clauses may be translated to the following
Hamiltonian, 

$$
H_{final}=\sum_{\alpha = 1}^m\ket{b_j^\alpha b_k^\alpha blj^\alpha}\bra{b_j^\alpha b_k^\alpha blj^\alpha}.
$$
A common approach to solve this type of constraint satisfaction problem is to use adiabatic quantum computations(AQC). One first prepares the ground states of an easy-to-solve Hamiltonian $H_{init}$. Next, one slowly evolves the Hamiltonian such that it adiabatically connects $H_{init}$ and $H_{final}$. In other words, the adiabatically evolved Hamiltonian reads
$$
H(s)=(1-s)H_{init}+sH_{final}, s \in [0,1],
$$
where $H_{init}$ is typically chosen to be a sum of one-qubit Hamiltonians $H_i$ acting on
the i-th qubit,
$$
H_{init}=\frac{1}{2}\sum_{i=1}^{n}h_i,\quad h_i=\begin{pmatrix}
    1 & -1 \\
   -1 & 1
\end{pmatrix}.
$$
The energy gap $\Delta E(s)$ between the instantaneous ground state and first-excited state of $H(s)$ will vary with time $s$
as shown in Fig~.\ref{dEdS}. Based on the well-established theoretical studies, it is also clear that the instantaneous energy gap of interest can get very small along the adiabatic path when the system size is large.

\FloatBarrier
\section{The imaginary time Control Strategy} \label{appstrategy}
For the imaginary-time control, in this letter, we mainly used two types of control strategy. The first one is the approximate bang-bang control introduced above that target the larger energy gap problem (type I). The second one is the modified version of the bang-bang strategy that targets the small energy gap problem (type II). We also provide the source codes of all simulations at \url{https://github.com/Kesson-C/QITC}, where you can find more details on the control strategy and control parameter settings.
\subsection{The imaginary time Control Strategy- Type I} \label{appsatistrategyI}
\begin{figure}[h]
\begin{center}
\includegraphics[width=0.9\textwidth]{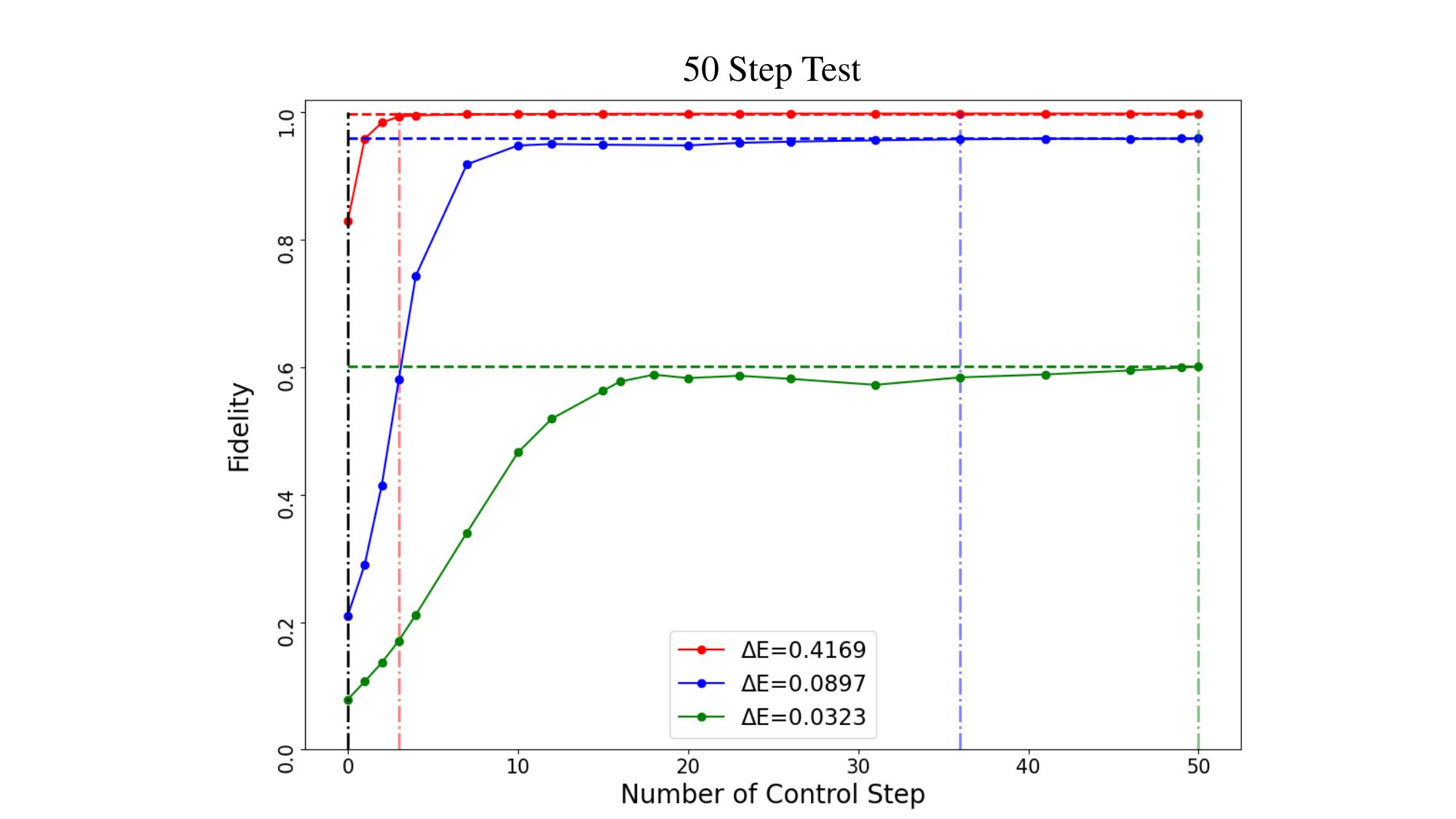}
\caption{The numerical control term truncation test result of 8 qubits HF molecule with different energy gap}\label{50step}
\end{center}
\end{figure}

For the type I control strategy, let us generalize the standard Lyapunov control such that it could work with the imaginary-time evolution. We redefine $T_k\equiv 2\expval{H_p}{\psi}\expval{H_d}{\psi}-\expval{\{H_p,H_d\}}{\psi}$ in this case. The $H_d$ related terms in $T_k(\tau)$ may entail lots of extra measurements if they cannot be obtained by measuring the Pauli terms appearing in $\expval{H_p}{\psi}$. To reduce the measurement cost and still maintain a powerful $H_d$ to provide an enhanced convergence, we propose the following strategy. We first decide if the state in the quantum circuit has high overlap with any eigenstate of $H_p$ or $H_d$ by checking the value of $T_k$, if $T_k < L$ then we do not apply any control pulses, otherwise we use a similar control strategy for the real-time case introduced above.
$$\beta_k(\tau)=\left\{
\begin{array}{rcl}
&\frac{2S}{1+e^{-\gamma T_k}}-S,&  (T_k\geq L)\\
&0,&  else
\end{array} \right.
$$
where L is some pre-defined threshold value. If the state is close to an eigenstate (i.e. $T_k(\tau)<L$), we should turn off the control field and let the system evolves under $H_p$ in the imaginary-time domain. This truncation can greatly reduce the measurement costs (for the implementation of the corresponding variational algorithm) in the region where the state will linearly converge to the eigenstate. Finally, we test how the truncation (i.e. setting $\beta_k(\tau)=0$ when $T_k(\tau)<L$) will affect the precision of the converged results given by truncating the control pulse in the middle of the convergence (Figure \ref{50step}). For a physical system with a large energy gap between the first-excited state and the ground state, it only requires a few control steps at the beginning then the result of convergence will be very close to the result without truncation. For systems with a small energy gap, the truncation will significantly affect the precision of the converged results.
\subsection{The imaginary time Control Strategy- Type II} \label{appsatistrategyII}
\begin{figure}[h]
\begin{center}
\includegraphics[width=0.9\textwidth]{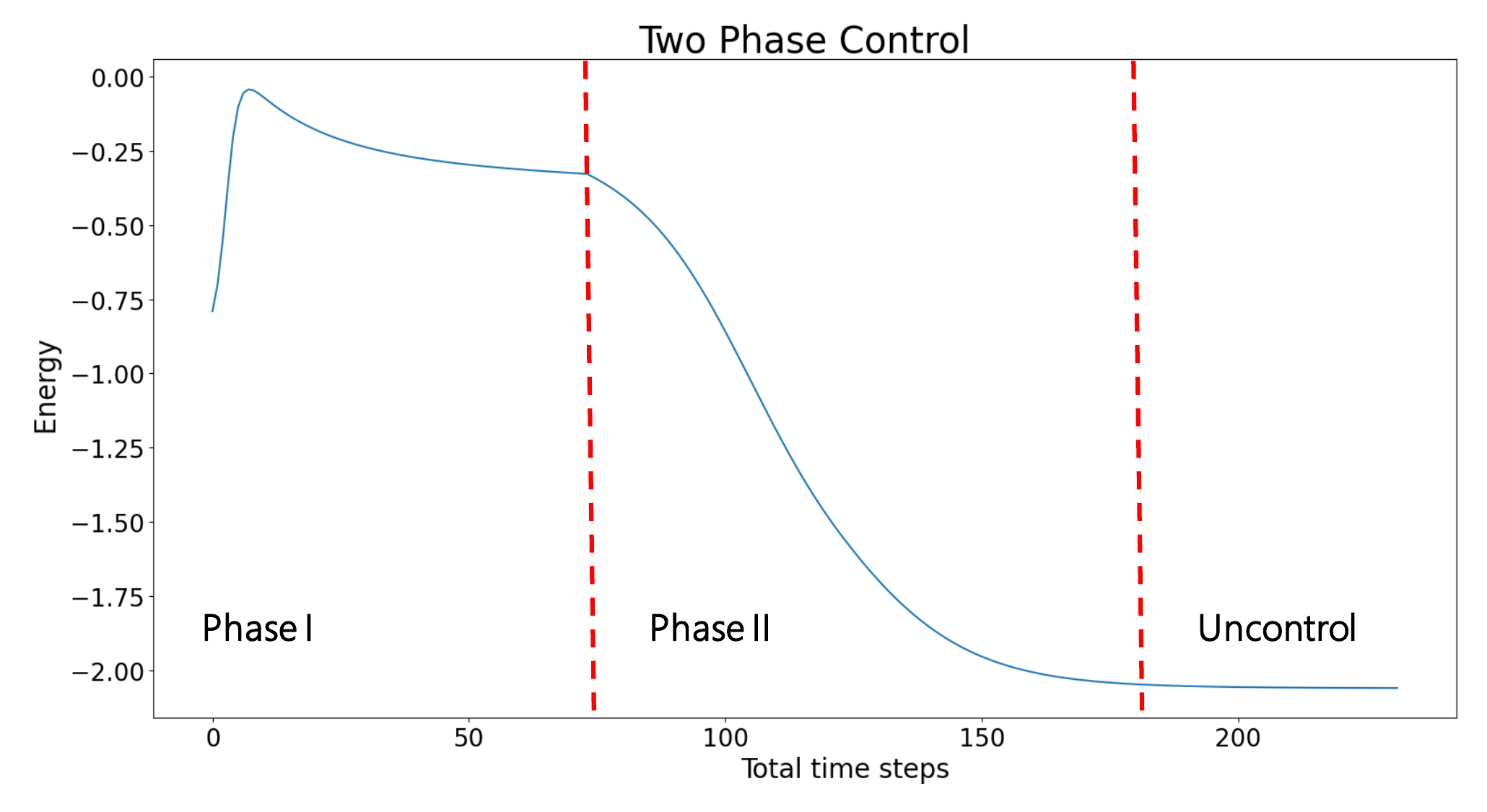}
\caption{Two-phase control example, the convergence will pass through three different phases. At phase I, the control will drift the state to a higher energy transition state. In phase II, the control will speed up the convergence to the ground state. In the final phase, the control will be closed to save computing resources.}\label{2phase}
\end{center}
\end{figure}

For the type II control strategy, we use a modified bang-bang control strategy that has two phases. The first phase of control will bypass the slow convergence region and the second phase will speed up the convergence, the demonstration of the phases is demonstrated in Figure \ref{2phase}. For the small energy gap problem, the type I control might drift the system to the low excited state and weaken the speed-up. To avoid this, we control the state to the transition state that contains high energy excited state and then to the ground state in two phases. Phase I of the control will lower the contribution of the low-energy states, the energy will go up and reach the equilibrium state consisting of high-energy excited states. Phase two is the traditional bang-bang control which is designed to lower the energy. The switch from phase I to phase II is decided by the energy changes (reach the equilibrium state), and the time-energy plot of the type two control is shown in Figure \ref{2phase}. In summary, we propose the following control strategy,
$$\beta_k(\tau)=\left\{
\begin{array}{rcl}
&K_1sgn(T_k(\tau)),&  (phase 1)\\
&-K_2sgn(T_k(\tau)),&  (phase 2)
\end{array} \right.
$$
where $K_1$ and $K_2$ are the strength of the control field and $sgn(T_k(\tau))$ is the sign of the $T_k(\tau)$, $\beta_k(\tau)$ will be set to zero if the sign alternating between positive and negative(reach the bound of the $H_d$), in this letter we set $K_1>>K_2$ to avoid convergence to the excited state.
In conclusion, the advantage of the type I control strategy has weak requirements for the selection of control Hamiltonian, but it may not provide exponential speed-up of the 1/$\Delta E$. The type II control strategy on the other hand can provide exponentially speed-up of the 1/$\Delta E$ but require stronger control Hamiltonian selection.

\FloatBarrier
\section{Molecular Hamiltonian} \label{appmolecularham}
\subsection{Hydrogen}
\begin{figure}[h]
\begin{center}

\includegraphics[width=0.8\textwidth]{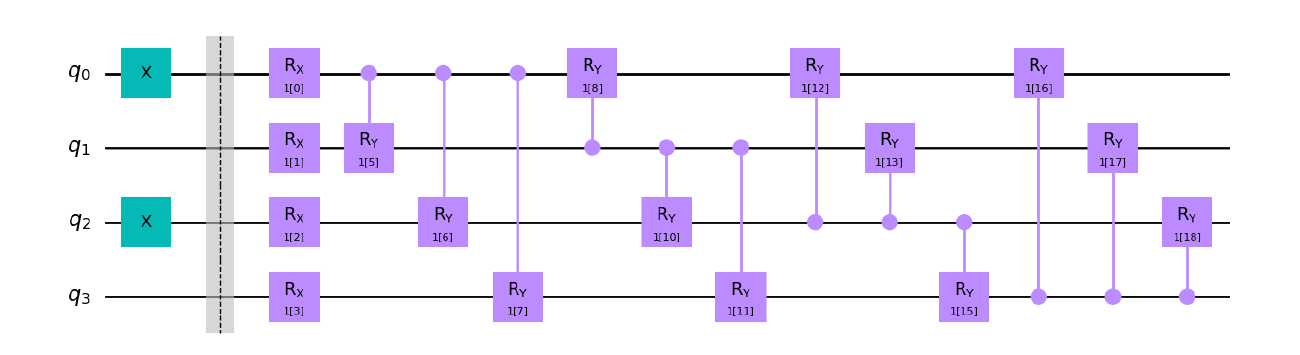}
\caption{The variational circuit of 4 qubit $H_2$ system}\label{cvqe}
\end{center}
\end{figure}

In our simulations, we consider the hydrogen molecule in the minimal STO-3G basis. Each
The hydrogen atom contributes a single 1S orbital. As a result of the spin, there are four spin-orbitals in total.
By using the function of the Qiskit, the qubit Hamiltonian for JW representation can be obtained. This 4-qubit Hamiltonian is given by:
\begin{align*}
H&=h_0I+h_1Z_0+h_2Z_1+h_3Z_2+h_4Z_3+h_5Z_0Z_1\\
&+h_6Z_0Z_2+h_7Z_0Z_3+h_8Z_1Z_2+h_9Z_1Z_3+h_{10}Z_2Z_3\\
&+h_{11}Y_0Y_1X_2X_3+h_{12}Y_0Y_1Y_2Y_3+h_{13}X_0X_1X_2X_3\\
&+h_{14}X_0X_1Y_2Y_3+h_{15}X_0X_1Y_2Y_3
\end{align*}
By using the function of the Qiskit, the 2-qubit Hamiltonian of the hydrogen molecule can be obtained from parity representation with Z2 symmetry reduction. This 2-qubit Hamiltonian is given by:
\begin{align*}
H&=h_0I+h_1Z_0+h_2Z_1+h_3Z_1Z_0+h_4X_1X_0
\end{align*}
And the circuit (Figure \ref{cvqe}) for Variational-based ansatz\cite{Sim2019} simulation used in the main text.
\par
To update the parameters of Variational-based ansatz, we calculate the gradient of the circuit by  using parameter shift rule\cite{Schuld2019} to obtain the numerical differential result of the circuit.

\subsection{Hydrogen Chain}
In our simulations, we consider the Hydrogen chain molecule in the minimal STO-3G basis. The Hydrogen Chain $H_n$ has n electrons and contributes n 1S orbital to the basis.

\subsection{Hydrogen Fluoride}
In our simulations, we consider the Hydrogen fluoride molecule in the minimal STO-3G basis. The Fluoride atom has 9 electrons, and so contributes a 1S, 2S, $2P_x$, $2P_y$, and $2P_z$ orbital to the basis, while the Hydrogen atom contributes a single 1S orbital. For the appendix B truncation test, by freezing the core two 1S orbitals, we can reduce the system from 12 orbitals with 10 electrons to 8 orbitals with 6 electrons on 2S and 2P orbitals. By using the Qiskit, the qubit Hamiltonian for Jordan-Wigner representation can be obtained. This 8-qubit Hamiltonian is given by 145 different Pauli terms and coefficients.

\subsection{Lithium Hydride}
In our simulations, we consider the Lithium hydride molecule in the minimal STO-3G basis. The Lithium atom has 3 electrons, and so contributes a 1S, 2S, $2P_x$, $2P_y$, and $2P_z$ orbital to the basis, while the Hydrogen atom contributes a single 1S orbital.